\documentclass[aps,prl,twocolumn,nofootinbib,notitlepage]{revtex4-2}

\usepackage{graphicx,color,tcolorbox}
\usepackage{amsmath,amsthm,amssymb,amsfonts,mathtools,latexsym,cancel,braket}
\usepackage[none]{hyphenat}

\usepackage[pdftex,colorlinks=true,linkcolor=Cerulean,citecolor=Cerulean,urlcolor=Cerulean]{hyperref}
\hypersetup{linkcolor=blue,citecolor=blue,urlcolor=blue}

\usepackage{array,booktabs,longtable,tabularx}
\usepackage{enumitem}
\usepackage{pifont}
\usepackage{changepage}
\usepackage{stackengine}
\stackMath

\DeclareFontFamily{OMS}{oasy}{\skewchar\font48}
\DeclareFontShape{OMS}{oasy}{m}{n}{
	<-5.5>		oasy5     <5.5-6.5>	oasy6
	<6.5-7.5>	oasy7     <7.5-8.5>	oasy8
	<8.5-9.5>	oasy9     <9.5->	oasy10
}{}
\DeclareFontShape{OMS}{oasy}{b}{n}{
	<-6>	oabsy5
	<6-8>	oabsy7
	<8->	oabsy10
}{}
\DeclareSymbolFont{oasy}{OMS}{oasy}{m}{n}
\SetSymbolFont{oasy}{bold}{OMS}{oasy}{b}{n}
\DeclareMathSymbol{\smallleftrightarrow}{\mathrel}{oasy}{"24}

\hypersetup
{
	pdfauthor={
		J. Enrique V\'azquez-Lozano (enrique.vazquez@unavarra.es),
		Victor Pacheco-Pe\~na (victor.pacheco-pena@newcastle.ac.uk) \&
		I\~nigo Liberal (inigo.liberal@unavarra.es)},
	pdfsubject={
		Department of Electrical, Electronic and Communications Engineering, Institute of Smart Cities (ISC), Universidad P\'ublica de Navarra (UPNA)
		School of Mathematics, Statistics and Physics, Newcastle University},
	pdftitle={
		Mirrors without spatial boundaries (Main Text)},
	pdfkeywords={temporal boundaries, time-varying media, non-Foster metamaterials, left-handed media, nanophotonics}
}

\begin{document}
\title{Mirrors without spatial boundaries}
\author{J. Enrique V\'azquez-Lozano}
\email{enrique.vazquez@unavarra.es}
\affiliation{Department of Electrical, Electronic and Communications Engineering, Institute of Smart Cities (ISC), Universidad P\'ublica de Navarra (UPNA), 31006 Pamplona, Spain}
\author{Victor Pacheco-Pe\~na}
\email{victor.pacheco-pena@newcastle.ac.uk}
\affiliation{School of Mathematics, Statistics and Physics, Newcastle University, Newcastle Upon Tyne, NE1 7RU, UK}
\author{I\~nigo Liberal}
\email{inigo.liberal@unavarra.es}
\affiliation{Department of Electrical, Electronic and Communications Engineering, Institute of Smart Cities (ISC), Universidad P\'ublica de Navarra (UPNA), 31006 Pamplona, Spain}

\date{\today}

\begin{abstract}
Mirrors are one of the most elementary and ubiquitous components of optical systems. They use a sharp refractive index contrast to provide the basic capability of reflecting light. Motivated by recent developments of photonic time-varying media, here we investigate the fundamental question on whether it is possible to have a mirror without any spatial boundary. In this vein, we first discuss how purely temporal mirrors are in general forbidden by the conservation of Minkowski momentum. However, we show that an exotic class of metamaterials, namely, temporal non-Foster left-handed media, exhibit anti-parallel Minkowski momentum and energy flow, thereby enabling mirrors without spatial boundaries. Upon this ground, we put forward some related photonic functionalities, including temporal cavities, pulse freezing, and frequency comb generators, which can be understood as the precursor of temporal lasers.
\end{abstract}

\maketitle
\sloppy

\section{Introduction}
\label{Sect.I}

In the last few years, {\em temporal and spatiotemporal metamaterials}~\cite{Yin2022,Galiffi2022,Caloz2020A,Caloz2020B,Caloz2022} have been proposed as promising platforms for the arbitrary control and manipulation of light-matter interactions in four-dimensions~\cite{Engheta2021,Yuan2022,Engheta2023}, namely, space $(x,y,z)$, and time, $t$. The underpinning idea consists in letting the constitutive electromagnetic (EM) parameters (namely, the permittivity, $\varepsilon$, and/or the permeability, $\mu$) to vary in time instead of (or in addition to) their conventional spatial dependence. Besides providing with just an additional degree of freedom, this approach has enabled a fundamentally different scenario wherein one may dynamically shape the optical properties of matter, thereby offering a greater~flexibility toward an enhanced control of EM waves~\cite{Taravati2020,Shaltout2019A,Pacheco-Pena2023A,Pacheco-Pena2024}. Moreover, it also offers a valuable alternative to overcome some practical limitations associated to the designs and fabrication of optical systems~\cite{Yoon2016,Hayran2023}.

Remarkably, both temporal and spatiotemporal modulation of the EM properties of media also entails important implications on the physics of the systems~\cite{Pacheco-Pena2022}, involving fundamental physical phenomena such as reciprocity~\cite{Sounas2017,Li2022}, causality~\cite{Solis2021A,Koutserimpas2024}, and energy balance~\cite{Hayran2022,Pendry2021}. Such stemming effects have been shown in a wide variety of platforms, including water~\cite{Bacot2016}, matter~\cite{Dong2024}, and EM waves from the millimetre to the optical regimes~\cite{Hadad2015,Martinez-Romero2016,Shaltout2019B,Pacheco-Pena2020A,Huidobro2021,Alex-Amor2023}. This has led to a wealth of novel applications as well as the extension of long-standing optical concepts to time-varying systems, including inverse prisms~\cite{Akbarzadeh2018}, temporal aiming~\cite{Pacheco-Pena2020B}, antireflection temporal coatings~\cite{Pacheco-Pena2020C}, filters~\cite{Ramaccia2021,Castaldi2021}, the temporal equivalent to the Brewster angle~\cite{Pacheco-Pena2021}, metasurface-based photonic time-crystals~\cite{Wang2023}, and the possibility for accumulating energy~\cite{Mirmoosa2019}. Recent studies have also theoretically demonstrated the potential of time-modulated media for controlling thermal radiation~\cite{Vazquez-Lozano2023A} and quantum optical features~\cite{Mendonca2000,Mendonca2005,Kort-Kamp2021,Vazquez-Lozano2023B,Liberal2023}, with remarkable experimental demonstrations of time-varying media at different ranges~\cite{Bacot2016,Tirole2022,Tirole2023,Moussa2023,Nicholls2017}.

Mirrors are an essential passive component across the entire realm of optics and photonics, spanning from the very free-space optical setups, to sophisticated nanophotonic systems, such as optical cavities, absorbers, or lasers~\cite{Born,Novotny}. However, despite extensive discussions on the temporal counterparts of elementary optical processes like transmission, refraction, and reflection~\cite{Morgenthaler1958,Fante1958,Xiao2014,Mendonca2002,Zurita-Sanchez2009,Plansinis2015,Ramaccia2020,Gratus2021,Mai2023,Mostafa2024,Ptitcyn2023A}, the notion of a totally reflective temporal boundary, capable of bringing forth a (perfect) temporal mirror, has thus far remained quite elusive. Indeed, owing to the fundamental differences between the spatial and temporal features on the conservation of Minkowski momentum in time-varying media~\cite{Ortega-Gomez2023}, a purely temporal mirror based on conventional temporal boundaries turns out to be forbidden. Such a prescription may also be understood by noting that the implementation of temporal mirrors in conventional temporal boundaries is physically unfeasible since the corresponding temporal Fresnel transmission coefficient for the forward wave cannot be totally suppressed~\cite{Morgenthaler1958,Fante1958,Xiao2014,Mendonca2002,Mai2023,Mostafa2024}.

Here, we propose the idea of {\em temporal mirrors without spatial boundaries} and theoretically demonstrate that they could be possible with the use of {\em temporal non-Foster left-handed media} (NF-LHM). It must be noted that this concept is essentially different from recent proposals on time-varying mirrors~\cite{Tirole2022,Tirole2022,Moussa2023}, where the properties of a spatial mirror are modulated at ultrafast time scales. Conversely, here we leverage recently introduced concepts of {\em temporal non-Foster metamaterials}~\cite{Hrabar2020,Hrabar2022,Ptitcyn2022,Kiasat2018,Pacheco-Pena2018,Pacheco-Pena2023B,Ptitcyn2023B}, and {\em temporal negative refraction}~\cite{Lasri2023}, with the concomitant notion of {\em non-Foster left-handed temporal boundaries} (NF-LHTBs), i.e., temporal boundaries between two media with opposite-sign refractive index. Upon this basis, we consider the scenario where a light pulse propagates through an unbounded medium undertaking a NF-LHTB, that is, a boundary where the refractive index of the entire space is rapidly changed from a positive to a negative value at a given time. Importantly, we assume that such a NF-LHTB occurs in a frequency range where both the permittivity and the permeability on each side of the temporal boundary (i.e., both for the positive and the negative refractive index) exhibit uniform bandwidth, wide enough so that both media can be effectively regarded as non-dispersive~\cite{Tretyakov2001,Engheta2005}. Despite the inherently dispersive character of such a system~\cite{Solis2021A,Koutserimpas2024,Hayran2022,Tretyakov2007,Solis2021B,Zhang2021}, its physical implementation could be carried out with the use of non-Foster elements~\cite{Hrabar2010A}, which have recently found a profitable niche of applicability in the context of (spatio)temporal metamaterials~\cite{Kiasat2018,Pacheco-Pena2018,Pacheco-Pena2023B,Ptitcyn2023B}. Finally, by considering different configurations of this class of temporal boundaries implemented with NF-LHM, we put forward some related photonic functionalities, such as {\em temporal cavities}, {\em freezed light pulses}, and {\em frequency-comb generators}, which can be envisioned as the precursors of {\em temporal lasers}.

\section{Spatial versus temporal mirrors}
\label{Sect.II}

\begin{figure*}[t!]
	\centering
	\includegraphics[width=0.95\linewidth]{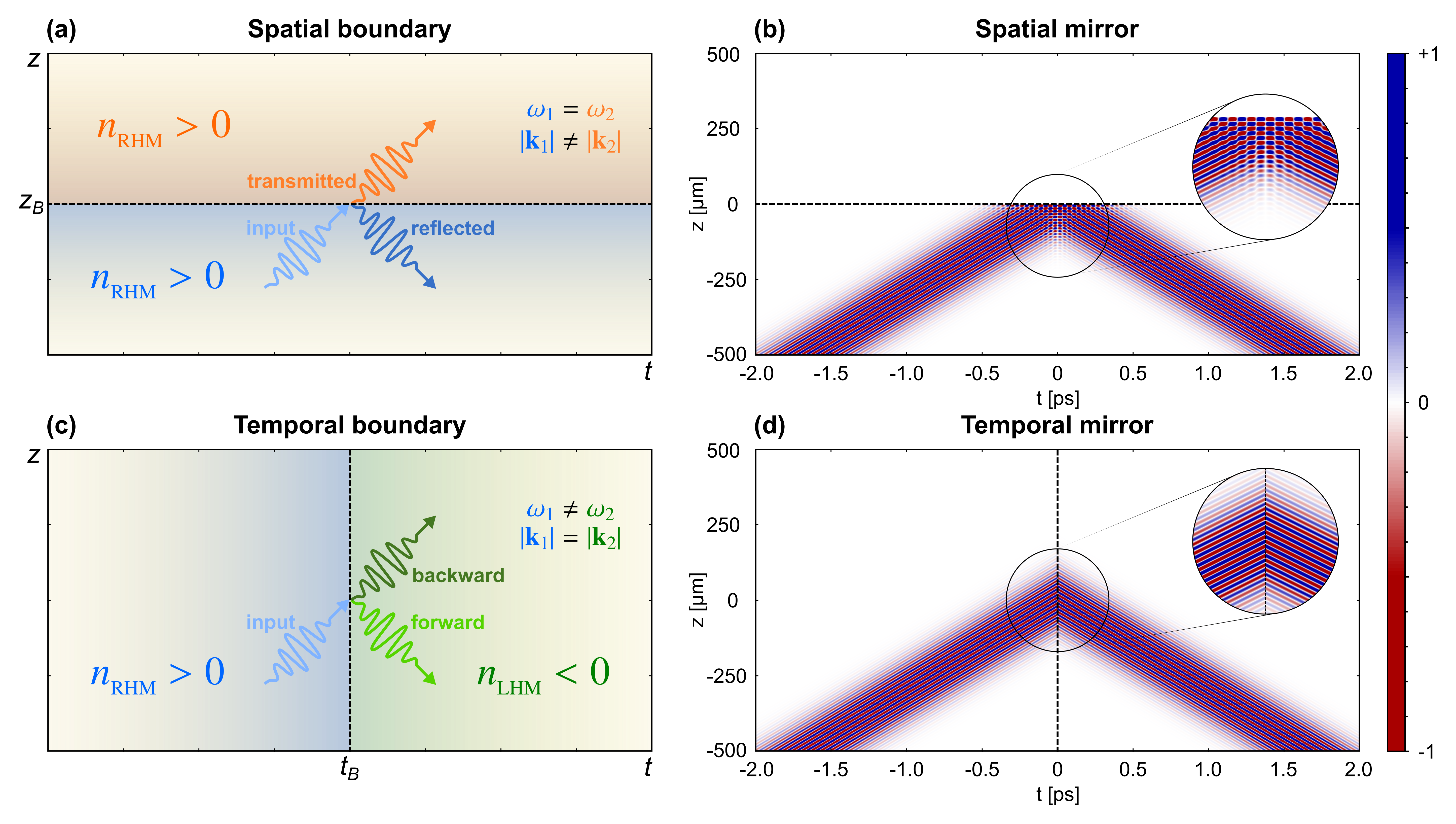}
	\caption{\textbf{Spatial versus temporal mirrors.} \textbf{(a)} Input, transmitted, and reflected EM wave propagating through a spatial boundary. \textbf{(b)} Electric field intensity of a Gaussian light pulse propagating through a totally reflecting spatial boundary. \textbf{(c)}~Input, forward, and backward EM wave propagating through a temporal boundary. \textbf{(d)} Electric field intensity of a Gaussian light pulse propagating through a totally reflecting non-Foster left-handed temporal boundary (NF-LHTB). The zoom-in view shows the phase switching associated to each transition.}
	\label{Fig.01}
\end{figure*}

The realization of a conventional mirror essentially relies on the existence of a spatial boundary separating two media with different constitutive parameters displaying a sharp refractive index contrast~\cite{Born,Novotny}. In general, when an EM wave encounters a spatial boundary, it is splitted into two contributions: one that keeps on moving forward (often referred to as transmitted or refracted wave), and another turning back on its propagation (referred to as reflected wave)~[see~\hyperref[Fig.01]{Fig.~1(a)}]. Thus, the goal of a mirror is to ideally reach total reflection and null transmission~[see~\hyperref[Fig.01]{Fig.~1(b)}]. Yet, on account of the fundamentally distinct characteristics between spatial and temporal boundaries for EM waves~\cite{Gratus2021,Mai2023,Mostafa2024}, the underlying conditions to get the temporal counterpart of a mirror are radically different from the spatial approach.

Qualitatively, the most evident difference between spatial and temporal boundaries roots on their own physical nature~\cite{Gratus2021,Mai2023,Mostafa2024}. Specifically, spatial boundaries are physical interfaces, not necessarily flat, that exist permanently for all times. Thus, the interface dividing the original space in two media sets down two different half-spaces with different material properties that separately host the input and reflected waves from the transmitted wave~[see~\hyperref[Fig.01]{Fig.~1(a)}]. By contrast, temporal boundaries, defined over a temporal plane, are ephemeral, so that, once produced are suddenly faded away. Hence, all the waves resulting from the interaction with a temporal boundary propagate in the same unbounded medium~[see~\hyperref[Fig.01]{Fig.~1(c)}]. Another qualitative difference arises on account of dimensional considerations. Spatial degrees of freedom allow one to define angles of incidence with respect to the interface, and, according to Snell's laws, so for the reflected and refracted (or transmitted) waves. Consequently, besides spatial mirrors enabled by high refractive index contrast between two media, there are purely geometrical configurations based on spatial boundaries that enable perfect reflection regardless the specific material properties, e.g., that leading to the phenomenon of {\em total internal reflection}~\cite{Born,Novotny}. By contrast, since temporal boundaries are, by construction, always orthogonal to the direction of propagation of the input wave, the notion of incidence angle, and hence the usual form of the Snell's laws as well as the critical angle become meaningless~\cite{Mendonca2002}. Interestingly, it has been shown that angles of incidence become relevant when using isotropic-to-anisotropic temporal boundaries~\cite{Pacheco-Pena2021}, and a generalization of Snell's in 4D has recently been reported~\cite{Pacheco-Pena2023A}. Therefore, assuming temporal transitions between linear, homogeneous, isotropic, non-dispersive, and lossless media, the temporal analogues to reflection and transmission simply consist of two EM waves, referred in the literature to as backward and forward waves~\cite{Morgenthaler1958,Fante1958,Xiao2014,Mendonca2002,Zurita-Sanchez2009,Plansinis2015,Ramaccia2020,Gratus2021,Mai2023,Mostafa2024,Ptitcyn2023A}, propagating in opposite directions to each other with respect to the incident input wave~[see~\hyperref[Fig.01]{Fig.~1(c)}]. This fact allows us to restrict the problem to only one spatial dimension~[see~\hyperref[Fig.01]{Fig.~1}].

In relation with the above qualitative considerations, there are further analytical differences between spatial and temporal boundaries that should be accounted for the realization of temporal mirrors~\cite{Gratus2021,Mai2023,Mostafa2024}. Particularly relevant is the fundamental asymmetry in the field boundary (or continuity) conditions for the spatial and temporal interfaces~\cite{Morgenthaler1958,Fante1958,Xiao2014,Mendonca2002}. Indeed, in the case of spatial boundaries, the field continuity conditions are imposed on the tangential components of the electric and magnetic fields, i.e., so that ${\bf E}_\parallel(z_0^-)={\bf E}_\parallel(z_0^+)$ and ${\bf H}_\parallel(z_0^-)={\bf H}_\parallel(z_0^+)$, where, without any loss of generality, it is assumed that the spatial boundary is set at $z=z_0$. On the other hand, for a temporal boundary, the field continuity conditions are imposed on the displacement and the magnetic induction fields~\cite{Morgenthaler1958,Fante1958,Xiao2014,Mendonca2002,Zurita-Sanchez2009,Plansinis2015,Ramaccia2020,Gratus2021,Mai2023,Mostafa2024}, i.e., so that ${\bf D}_\parallel(t_0^-)={\bf D}_\parallel(t_0^+)$ and ${\bf B}_\parallel(t_0^-)={\bf B}_\parallel(t_0^+)$, assuming that the temporal boundary is set at $t=t_0$. Here it is worth emphasizing that, since the EM fields are vectors whose components are characterized in terms of their spatial components, and hence always orthogonal to the temporal axis, the continuity conditions for a temporal boundary straightforwardly apply to the whole ${\bf D}$ and ${\bf B}$ fields, thus not needing to distinguish between tangential and normal components. This could be then considered as the closest equivalent to the case of a spatial boundary under normal incidence. Notice however that, in the case of either spatially or temporally dispersive boundaries, the corresponding continuity conditions differ from the above, in such a way that the entire vector fields are to be continuous across the temporal interface~\cite{Hayran2022,Solis2021B,Zhang2021,Alvarez2020}. At any rate, both conditions, either those involving ${\bf E}$~and~${\bf H}$ for spatial boundaries, or ${\bf D}$ and ${\bf B}$ for temporal boundaries, are established to ensure the charge and the total magnetic flux conservation~\cite{Morgenthaler1958,Mostafa2024}.

Elaborating further upon this fundamental ground, it must be noted that spatial and temporal boundaries are ultimately underpinned respectively by the energy and the (Minkowski) momentum conservation laws~\cite{Mai2023,Mostafa2024,Ptitcyn2023A,Ortega-Gomez2023}. Such a correspondence can be readily understood by noticing that for spatial and temporal interfaces, the corresponding change in the speed of light associated with each transition, is to be respectively accomplished by supplying either momentum or energy. Whereas momentum change is automatically undertaken by the stationary perturbation of the refractive index, without expending energy, the energy transfer needed to shape the light in temporal boundaries is to be externally sourced. From another perspective, such a correspondence between the transformations in spatial/temporal boundaries with the energy/momentum conservation may also be neatly evinced from the relationship between spatial and temporal features either with the wavevector, ${\bf k}$, and the frequency,~$\omega$, which, in other contexts have proven to be clearly related with momentum, $\hbar{\bf k}$, and energy, $\hbar\omega$~\cite{Hayran2022,Sakurai,Shankar,Cohen-Tannoudji}. Likewise, on account of Noether's theorem~\cite{Kosmann-Schwarzbach,Banados2016}, the prevalence of conserved quantities is closely associated with the presence of continuous symmetries~\cite{Fushchich}. Accordingly, inasmuch as spatial boundaries break spatial translational symmetries, not affecting the temporal dimension, they bring about a transformation of the Minkowski momentum at the same time that preserve the energy conservation. Conversely, temporal boundaries preserve Minkowski momentum while changing the energy. So, for spatial boundaries, the input (${\rm I}$), reflected (${\rm R}$), and transmitted (${\rm T}$) waves fulfill the following relations for the frequency and the~wavenumber (i.e., the modulus of the wavevector): $\omega_{\rm I}=\omega_{\rm R}=\omega_{\rm T}$, $k_{\rm I}=-k_{\rm R}\neq k_{\rm T}$. Similarly, for temporal boundaries, the frequency and wavenumber relations for the input, backward (${\rm BW}$), and forward (${\rm FW}$) waves read as: $\omega_{\rm I}\neq \omega_{\rm BW}=\omega_{\rm FW}$, $k_{\rm I}=-k_{\rm BW}= k_{\rm FW}$.

The integration of the above frequency-wavenumber relations into the corresponding field continuity conditions associated with the spatial and temporal boundaries leads to the corresponding Fresnel reflection and transmission coefficients. Specifically, in the case of a spatial boundary (SB), assuming a normal wave incidence configuration, from the electric field ratios it is straightforward to show that the conventional spatial Fresnel reflection and transmission coefficients read as~\cite{Born,Novotny,Pozar}:
\begin{subequations}
\begin{align}
r_{\rm SB}=\frac{E_{\rm R}}{E_{\rm I}}=\frac{Z_2-Z_1}{Z_2+Z_1},\\
t_{\rm SB}=\frac{E_{\rm T}}{E_{\rm I}}=\frac{2Z_2}{Z_2+Z_1},
\end{align}
\end{subequations}
where $Z_i=\sqrt{\mu_i/\varepsilon_i}$ stands for the wave impedance in each medium conforming the spatial boundary, assumed to be non-dispersive, and hence lossless. In turn, these expressions yield the EM energy conservation for the input, transmitted, and reflected contributions:
\begin{equation}
S_1=Z_1^{-1}\left|{\bf E}_{\rm I}\right|^2=Z_2^{-1}\left|{\bf E}_{\rm T}\right|^2+Z_1^{-1}\left|{\bf E}_{\rm R}\right|^2=S_2,
\end{equation}
where $S\equiv\hat{z}\cdot[{\bf E}\times{\bf H}]$ is the Poynting vector along the propagation axis, describing the EM power intensity, with $H=Z^{-1}E$. This can be expressed through the balance equation:
\begin{equation}
T_{\rm SB}+R_{\rm SB}=1,
\end{equation}
where $T_{\rm SB}=[Z_1/Z_2]|t_{\rm SB}|^2$ and $R_{\rm SB}=|r_{\rm SB}|^2$ are, respectively, the transmittance and the reflectance normalized with respect to the EM power of the input wave. Here it is worth remarking that the above balance equation is referred to the EM power before and after the spatial boundary, which should not be confused with the balance equation for the EM fields, $t_{\rm SB}-r_{\rm SB}=1$, accounting for the EM intensity at each side of the spatial~boundary.

On the other side, for the temporal boundary (TB), the corresponding continuity conditions for the ${\bf D}$ and ${\bf B}$ fields, along with the corresponding frequency-wavenumber relations, bring about the temporal counterparts of the Fresnel reflection and transmission coefficients~\cite{Xiao2014,Mendonca2002,Mai2023,Mostafa2024}~(see~\hyperref[AppendixA]{\em Appendix~A} for further details on the derivation):
\begin{subequations}
\begin{align}
r_{\rm TB}=\frac{E_{\rm R}}{E_{\rm I}}=\frac{\varepsilon_1}{\varepsilon_2}\left[\frac{Z_2-Z_1}{2Z_2}\right],\\
t_{\rm TB}=\frac{E_{\rm T}}{E_{\rm I}}=\frac{\varepsilon_1}{\varepsilon_2}\left[\frac{Z_2+Z_1}{2Z_2}\right].
\end{align}
\end{subequations}
As anticipated, temporal boundaries do not conserve the EM energy, but they actually preserve the Minkowski momentum for the input, forward, and backward contributions:
\begin{equation}
P_1=Z_1\left|{\bf D}_{\rm I}\right|^2=Z_2\left|{\bf D}_{\rm FW}\right|^2-Z_2\left|{\bf D}_{\rm BW}\right|^2=P_2,
\end{equation}
where $P\equiv\hat{z}\cdot[{\bf D}\times{\bf B}]$ stands for the Minkowski momentum along the propagation axis, with $B=ZD$. This can be compactly expressed by means of a balance equation in terms of the temporal Fresnel reflection and transmission coefficients:
\begin{equation}
T_{\rm TB}-R_{\rm TB}=1,
\end{equation}
being $T_{\rm TB}=(\varepsilon_2/\varepsilon_1)^2[Z_2/Z_1]|t_{\rm TB}|^2$ and $R_{\rm TB}=(\varepsilon_2/\varepsilon_1)^2[Z_2/Z_1]|r_{\rm TB}|^2$ the temporal counterpart to the transmittance and the reflectance, respectively, in this case, normalized with respect to the Minkowski momentum associated to the input wave. Again, it is worth remarking that the above balance equation is meant for the Minkowski momentum before and after the temporal boundary, that should not be confused with the balance for the EM fields, $t_{\rm TB}+r_{\rm TB}=\varepsilon_1/\varepsilon_2$, which, similar to the spatial case, account for the EM intensity at each side of the temporal boundary.

\begin{figure}[t!]
	\centering
	\includegraphics[width=1\linewidth]{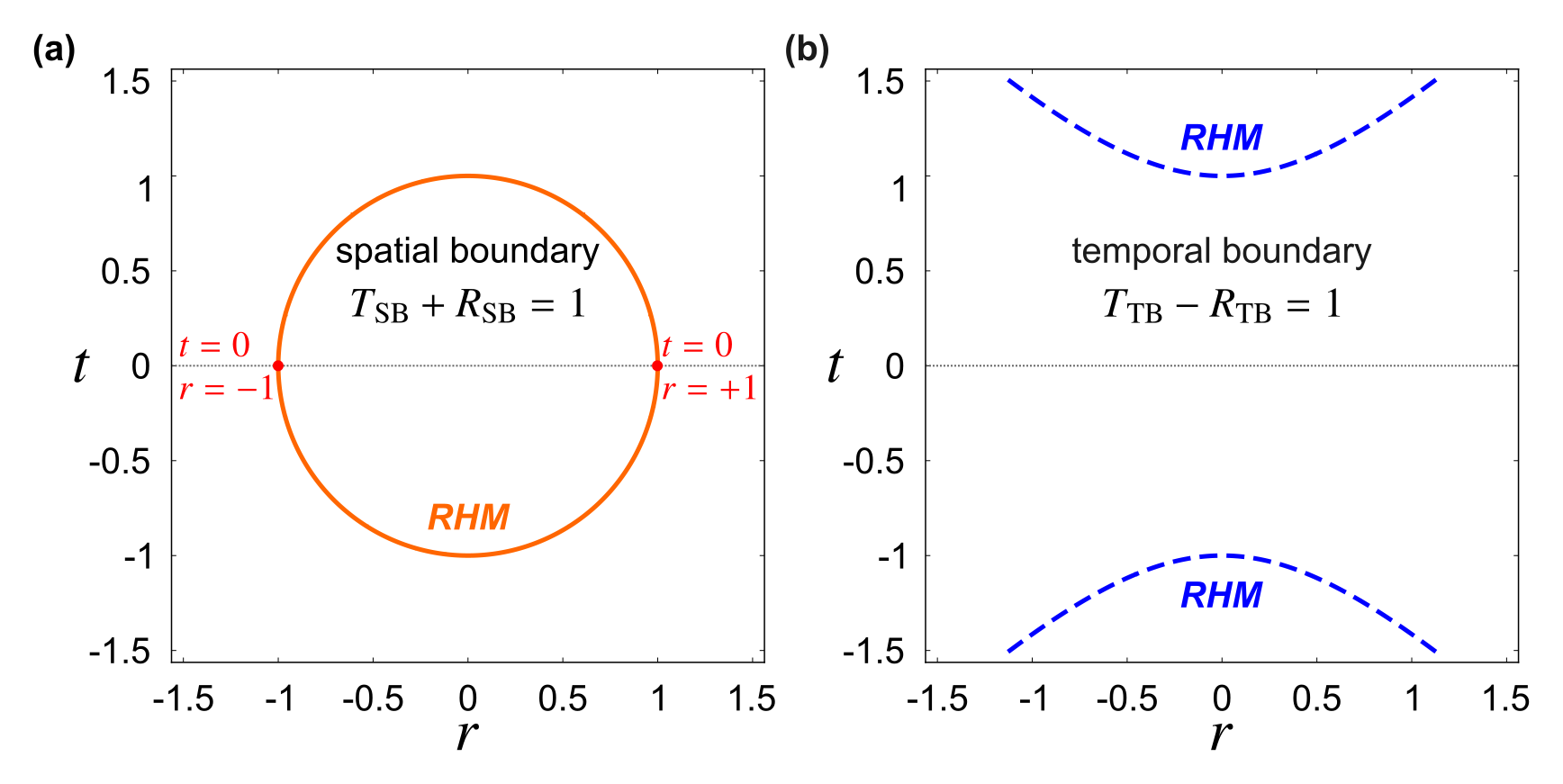}
	\caption{\textbf{Balance equation for the transmittance and reflectance of spatial and temporal boundaries.} \textbf{(a)} The transmittance-reflectance balance equation of spatial boundaries is tied to the EM energy conservation, and displays a simply connected (unit-circle) topological structure. Accordingly, the realization of spatial mirrors simply rely on the condition of unit reflectance and null transmittance. \textbf{(b)} The transmittance-reflectance balance equation of temporal boundaries is tied to the Minkowski momentum conservation, and displays a non-simply connected (hyperbolic) topological structure. Accordingly, temporal mirrors are, a priori, theoretically forbidden, as there are no solutions for $t=0$.}
	\label{Fig.02}
\end{figure}

The EM energy and Minkowski momentum conservation, and their associated balance equations, respectively related with the spatial and temporal boundaries, enclose a crucial, though subtle, implication as for the realization of (perfect) mirrors. Indeed, for spatial boundaries, the EM energy conservation leads to a simply connected (unit-circle) transformation, so that the condition of unit reflectance and null transmittance, i.e., $R_{\rm SB}=1$ and $T_{\rm SB}=0$, is mathematically attainable~[see~\hyperref[Fig.02]{Fig.~2(a)}]. In the case of temporal boundaries, however, the conservation of the Minkowski momentum leads to a hyperbolic transformation with unconstrained energy production that prevents the realization of temporal mirrors. Indeed, such a limitation can be clearly drawn from the balance equation for the temporal boundaries, just by noticing that $T_{\rm TB}=0$ and $R_{\rm TB}=1$ yield a mathematical inconsistency~[see~\hyperref[Fig.02]{Fig.~2(b)}]. Notwithstanding the foregoing, it is still possible to theoretically overcome this constraint for the realization of temporal mirrors by reconsidering the physical meaning of transmittance and reflectance and their relationship with the resulting forward and backward waves generated at temporal boundaries. To this aim, in the following section we put forward a configuration based on temporal NF-LHM enabling the realization of NF-LHTBs, namely, temporal boundaries switching the handedness of the media, from positive to negative refractive index~[see~\hyperref[Fig.01]{Fig.~1(c)}].

\section{Temporal mirrors enabled by non-Foster left-handed temporal boundaries}
\label{Sect.III}

On the basis of the above discussion, our proposal of temporal mirrors without spatial boundaries relies on the realization of NF-LHTBs, that is, the simultaneous implementation of time-varying and left-handed media in a frequency range where dispersion can be disregarded.

It should be noted that, when considering optical waves propagating through LHTBs, the usual notions of forward and backward waves associated to conventional right-handed temporal boundaries, namely, temporal boundaries between right-handed media (RHM), i.e., media with different but positive refractive index, turn out to be switched~[see~\hyperref[Fig.01]{Fig.~1(c)}]. Of course, this distinction lies on how we physically characterize the EM wave propagation. Thus, by assuming that the EM modes can be characterized in terms of the wavevector ${\bf k}$ the corresponding Poynting vector and group velocity along the propagation in RHM are so that:
\begin{subequations}
\begin{align}
{\bf S}^{+k}>0, & & {\bf S}^{-k}<0,\\
v_{\rm g}^{+k}>0, & & v_{\rm g}^{-k}<0.
\end{align}
\end{subequations}

By contrast, according to our proposal, the EM wave after the temporal boundary propagates through LHM, so that the phase evolution (characterized in terms of the phase velocity, ) and the energy flow (which can be described by the group velocity, $v_{\rm g}$, or, alternatively, by means of the Poynting vector, ${\bf S}\equiv{\bf E}\times{\bf H}$) are anti-parallel~\cite{Ziolkowski2001,Veselago2006}. Therefore, in such a case, it follows~that:
\begin{subequations}
\begin{align}
{\bf S}^{+k}<0, & & {\bf S}^{-k}>0,\\
v_{\rm g}^{+k}<0, & & v_{\rm g}^{-k}>0.
\end{align}
\end{subequations}
This feature, usually referred to as the {\em backward propagation}, ultimately lies on the phenomenon of negative refraction~\cite{Pendry2004,Pendry2008}, and is yielded by a medium with both the permittivity and permeability negative, so that the refractive index displays the negative sign, $n=\pm\sqrt{\varepsilon\mu}$.

\begin{figure}[t!]
	\centering
	\includegraphics[width=1\linewidth]{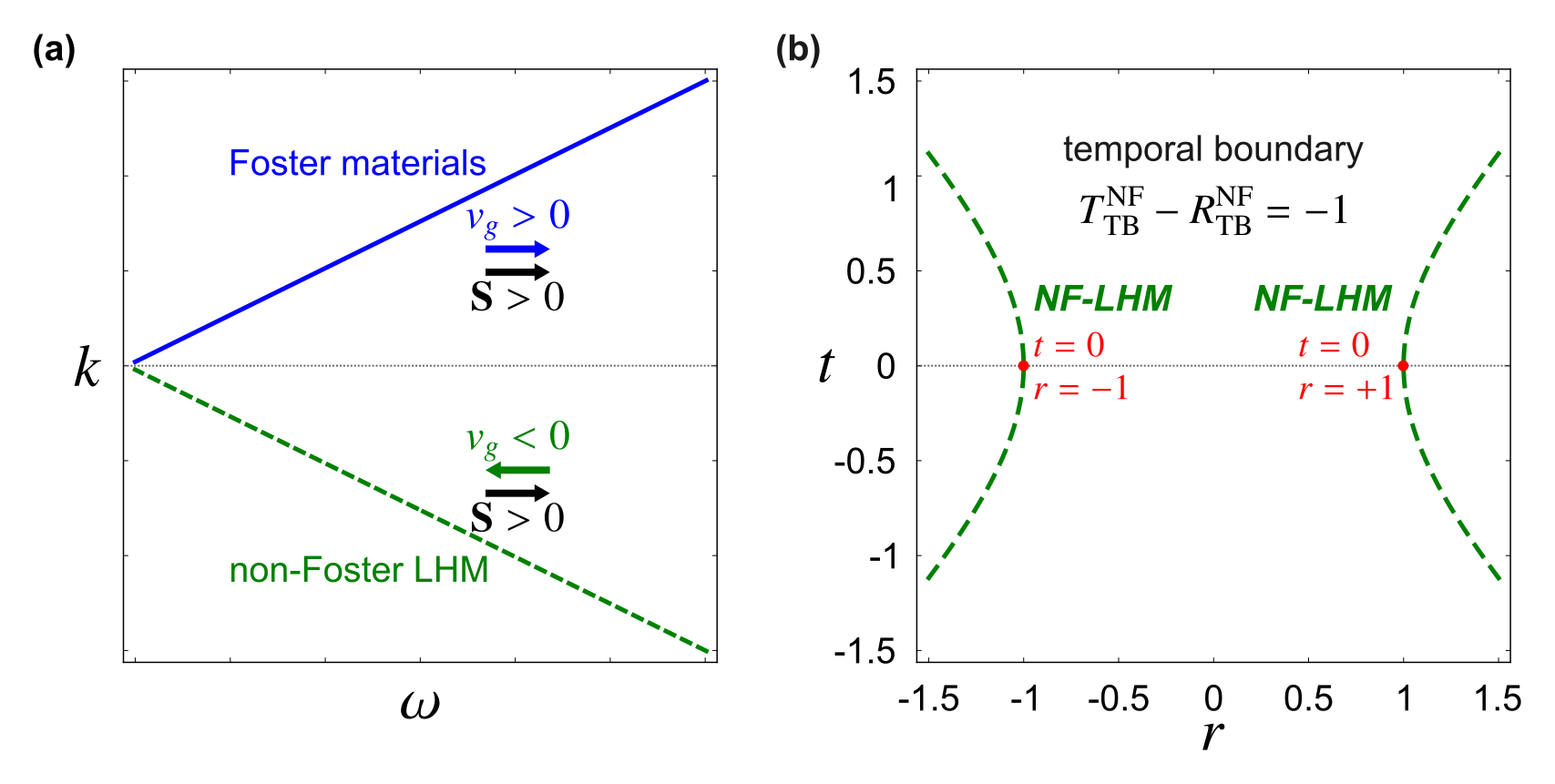}
	\caption{\textbf{Theoretical characterization of temporal mirrors enabled by temporal non-Foster LHM.} \textbf{(a)} Dispersion diagram for Foster materials (solid) and non-Foster LHM (dashed), highlighting, respectively, the parallel and antiparallel character of the Poynting vector and group velocity of EM waves propagating~forward. \textbf{(b)}~Transmittance-reflectance balance equation for NF-LHTBs underscoring the theoretical condition for the realization of temporal mirrors (red spots).}
	\label{Fig.03}
\end{figure}

From these relations, it is clear that, in RHM, the wave propagation is so that the wavevector, the Poynting vector, and the group velocity are parallel, pointing toward the same direction. In LHM, however, the Poynting vector and group velocity are parallel to each other, but anti-parallel with respect to the wavevector, whose direction is related with the phase velocity. In any case, both the Poynting vector and the group velocity determine the energy flow~\cite{Caloz,Engheta}. This is actually the customary behavior of the so-called Foster materials~[see~\hyperref[Fig.03]{Fig.~3(a)}], namely, those fulfilling a relation analogue to the {\em Foster reactance theorem} in circuit theory~\cite{Foster1924}, according to which $\partial_\omega[\omega\varepsilon(\omega)]>0$ and $\partial_\omega[\omega\mu(\omega)]>0$. This theorem is always satisfied for RHM, even under the approximation where material dispersion is disregarded. However, in the case of LHM, the above relations for the wavevector, the Poynting vector, and the group velocity, necessarily requires the consideration of material dispersion, a condition that is inherently prescribed by the Foster reactance theorem. In fact, in non-dispersive media, group velocity and phase velocity are identical to each other, and hence, they should necessarily stand with the same sign~\cite{Dolling2006}, so that the above relations between the group velocity and the wavevector, i.e., $v_{\rm g}^{\pm k}\lessgtr
0$, turns out to be attainable for dispersive LHM.

Notwithstanding the foregoing, as in many of the previous studies on time-varying photonics~\cite{Morgenthaler1958,Fante1958,Xiao2014,Mendonca2002,Zurita-Sanchez2009,Plansinis2015,Ramaccia2020,Gratus2021,Mai2023,Mostafa2024,Ptitcyn2023A}, in this first theoretical approach for the realization of temporal mirrors without spatial boundaries, for simplicity, we consider a rather idealized scenario disregarding dispersive material features. This assumption is to be carefully undertaken, as both~time-varying and left-handed media are inherently~and strongly dispersive photonic platforms~\cite{Solis2021A,Koutserimpas2024,Hayran2022,Tretyakov2001,Engheta2005,Tretyakov2007,Solis2021B,Zhang2021}. In fact, according to the Kramers-Kronig relations, underpinning the principle of causality, it is impossible to realize a passive optical system whose constitutive parameters are simultaneously negative and real (i.e.,~lossless or gainless) in an arbitrarily large bandwidth~\cite{Ziolkowski2001,Veselago2006}. Likewise, within the context of time-varying media, dispersion is closely related with the causal response function of the material to an external bias~\cite{Ptitcyn2023A}, thereby preventing it to be instantaneous. Consequently, the effective implication of considering non-dispersive media is the limitation of the operative bandwidth where both the permittivity and the permeability may exhibit an uniform behavior before and after the temporal boundary, which should be wide enough so that both RHM and LHM can be effectively regarded as non-dispersive.

In order to overcome this constraint, here we consider the inclusion of additional active components commonly referred to as {\em non-Foster elements}~\cite{Hrabar2010A}, which provide with a suitable pathway to surpass the aforementioned Foster reactance theorem~\cite{Foster1924}, thus allowing for the fulfillment of the relations $\partial_\omega[\omega\varepsilon(\omega)]<0$ and $\partial_\omega[\omega\mu(\omega)]<0$. Despite their seemingly unstable and counter-intuitive character, these kind of active elements preserve the causality, which is justified on account of the limitation of the bandwidth in realistic situations~\cite{Ugarte-Munoz2012,Saadat2012}. Notably, since the early proposal and developments of negative impedance converters~\cite{Linvill1953,Gonzalez-Posadas2010}, non-Foster elements have been extensively investigated across a wide~range of platforms and metamaterial configurations~\cite{Hrabar2010B,Barbuto2013,Hrabar2013A}. On this basis, negative capacitors~\cite{Hrabar2011,Okorn2017} and inductors~\cite{Zanic2021} loaded in transmission lines have been demonstrated experimentally. Moreover, practical applications of non-Foster elements include broadband matching of small antennas~\cite{Albarracin-Vargas2016,Albarracin-Vargas2023}, recently achieving a 198\% bandwidth~\cite{Wang2024}, and transmission lines emulating moving media~\cite{Vehmas2014}. Although NF-LHM have not been experimentally demonstrated yet, all-negative configurations are known to be stable, and RLC tanks with all-negative elements have been experimentally~demonstrated~\cite{Hrabar2013B}. Recently, time-varying media~\cite{Hrabar2020,Hrabar2022,Ptitcyn2022,Kiasat2018,Pacheco-Pena2018,Pacheco-Pena2023B,Ptitcyn2023B} and dispersion engineering~\cite{Qin2023} are being proposed as an unconditionally stable route toward the realization of non-Foster elements~\cite{Hrabar2018}.

Yet, regarding the realization of temporal mirrors, the paramount consequence of NF-LHM is the lifting of fundamental asymmetries in the characterization of the EM wave propagation in terms of the energy transport. Specifically, while in Foster materials, the energy flow can be indistinctly described either by the group velocity or the Poynting vector, in the case of non-Foster materials, it is exclusively determined by the group velocity, which is anti-parallel to the Minkowski momentum, and hence to the Poynting vector~[see~\hyperref[Fig.03]{Fig.~3(a)}]:
\begin{subequations}
\begin{align}
{\bf S}^{+k}<0, & & {\bf S}^{-k}>0,\\
v_{\rm g}^{+k}>0, & & v_{\rm g}^{-k}<0.
\end{align}
\end{subequations}
Accordingly, phase and group velocity are now parallel, pointing toward the same direction, thereby enabling the realization of LHTBs in non-dispersive media~\cite{Dolling2006,Hrabar2018}, namely, what we refer to as NF-LHTBs. Furthermore, from this characterization and the Minkowski momentum conservation, it can be demonstrated that~(see~\hyperref[AppendixB]{\em Appendix~B} for further details):
\begin{equation}
T_{\rm TB}^{\rm NF}-R_{\rm TB}^{\rm NF}=-1.
\end{equation}
In contrast with the previous expression for conventional temporal boundaries between RHM, this latter transmittance-reflectance balance equation for NF-LHTBs, i.e., for temporal boundaries between RHM and NF-LHM, neatly allows for the condition of a (perfect) temporal mirror, namely, $T_{\rm TB}^{\rm NF}=0$ and $R_{\rm TB}^{\rm NF}=1$~[see~\hyperref[Fig.03]{Fig.~3(b)}]. Nonetheless, it is worth emphasizing that in this case the transmittance and reflectance coefficients are respectively associated to the BW and FW contributions~[see~\hyperref[Fig.01]{Fig.~1(c)}], and the corresponding temporal Fresnel reflection 
and transmission coefficients appear to be interchanged with respect to the previous case:
\begin{subequations}
\begin{align}
r_{\rm TB}^{\rm NF}=\frac{\varepsilon_1}{\varepsilon_2}\left[\frac{Z_2+Z_1}{2Z_2}\right]=t_{\rm TB},\\
t_{\rm TB}^{\rm NF}=\frac{\varepsilon_1}{\varepsilon_2}\left[\frac{Z_2-Z_1}{2Z_2}\right]=r_{\rm TB}.
\end{align}
\end{subequations}
From the above considerations, our proposal of {\em temporal mirrors without spatial boundaries} straightforwardly results from the realization of NF-LHTBs between RHM and NF-LHM. To illustrate such a phenomenon we consider a scenario as that depicted in~\hyperref[Fig.01]{Fig.~1(c)}, where an input light wave that initially propagates in an~unbounded medium with positive refractive index $n_1>1$ undergoes a NF-LHTB so that the refractive index is rapidly switched to $n_2<0$. Specifically, in the numerical depiction shown in~\hyperref[Fig.01]{Fig.~1(d)}, the input signal is assumed to be a Gaussian pulse, linearly polarized on the $x$-axis,~which propagates along the $z$-axis, so that ${\bf E}_{\rm input}\equiv{\bf E}_1(z,t)=\cos{(\varphi_1)}e^{-[\varphi_1^2/(2\omega^2T^2)]}\hat{\bf x}$, with $\varphi_{\rm input}\equiv\varphi_1=k_1z-\omega t$, where $k_1=\omega/v_1$, and $v_1=c/n_1$, respectively stand for the phase, the wavenumber,~and the phase velocity, which, for non-dispersive media, coincides with the group velocity. Accordingly, the magnetic field is ${\bf H}_{\rm input}\equiv{\bf H}_1(z,t)=[Z_1]^{-1}E_1(z,t)\hat{\bf y}$. After the NF-LHTB, which, for simplicity and without any loss of generality, is assumed to occur at $t=t_{\rm B}=0$, the EM field experiences a change so that ${\bf E}_{\rm output}\equiv{\bf E}_2(z,t)=r_{\rm TB}^{\rm NF}[\cos{(\varphi_2)}e^{-[\varphi_2^2/(2\omega^2T^2)]}]\hat{\bf x}$, where the phase changes to $\varphi_{\rm output}\equiv\varphi_2=\varphi_1+\omega t[v_1-v_2]/v_1=\varphi_1+\omega t[\Delta n/n_2]$, with $\Delta n=n_2-n_1$. Notice that we have implicitly imposed extreme conditions for the Fresnel coefficients, so that the NF-LHTB only produces forward output waves~[see~\hyperref[Fig.01]{Fig.~1(c)}]. Specifically, it has been considered a switching from a vacuum, i.e., $\varepsilon_1=1$, $\mu_1=1$ ($n_1=1$, $Z_1=1$), to a NF-LHM defined by the constitutive parameters $\varepsilon_2=-1$, $\mu_2=-1$ ($n_2=-1$, $Z_2=1$), thus fulfilling the impedance-matching condition, $Z_1=Z_2$. Therefore,~the~corresponding temporal Fresnel coefficients are $r_{\rm TB}^{\rm NF}=-1$ and $t_{\rm TB}^{\rm NF}=0$. This simple configuration allows us to show that the handedness of the time interface influences on both the amplitude of the output field, introducing a modulation by means of the temporal Fresnel reflection coefficient $r_{\rm TB}^{\rm NF}<0$, and the phase, ultimately determining the direction of propagation by means of the refractive index contrast, $\Delta n=n_2-n_1$, being negative for NF-LHTBs. Upon this ground, in~\hyperref[Fig.01]{Fig.~1(d)} we show the numerical depiction of the electric field intensity evolution with $\omega=100/2\pi$~THz and $T=200$~fs, thereby showing the behavior of a temporal mirror brought about by a NF-LHTB. As previously anticipated, since the horizontal axis represents the time, the tilted depiction of the field evolution means an unidirectional spatial propagation, in this case along the $z$-axis, with the angle denoting the wave velocity. Bearing this in mind,~\hyperref[Fig.01]{Fig.~1(d)} shows that the NF-LHTB may affect to both the velocity and the direction of propagation. Nonetheless, since the transition showcased is to a NF-LHM with $n=-1$, the modulus of the velocity remains unchanged, and hence the tilting angle describing the time evolution just is reversed. Furthermore, the zoom-in views underscore a subtle, though distinctive, feature of wave reflection processes involving the phase. Specifically, just at the rim of the NF-LHTB, it can be seen a $\pi$-shift onto the phase, which directly translates into an amplitude reversal~\cite{Xiao2014,Mendonca2002}.

\section{Stemming functionalities enabled by non-Foster left-handed temporal boundaries}
\label{Sect.IV}

Beyond the temporal mirrors, the realization of NF-LHTBs paves the way for the proposal and development of further fascinating photonic functionalities. Either devising temporal analogues that may reproduce optical phenomena so far carried out just by spatial boundaries, or looking into unique effects empowered by the temporal dimension, in this section we put forward three distinctive examples of potential applications based on this class of temporal boundaries: a temporal cavity~[see~\hyperref[Fig.04]{Fig.~4}], a configuration enabling to freeze a light pulse~[see~\hyperref[Fig.05]{Fig.~5}], and a frequency-comb generator~[see~\hyperref[Fig.06]{Fig.~6}].

\subsection{Temporal cavity}
\label{Sect.IV.a}

\begin{figure*}[t!]
	\centering
	\includegraphics[width=0.925\linewidth]{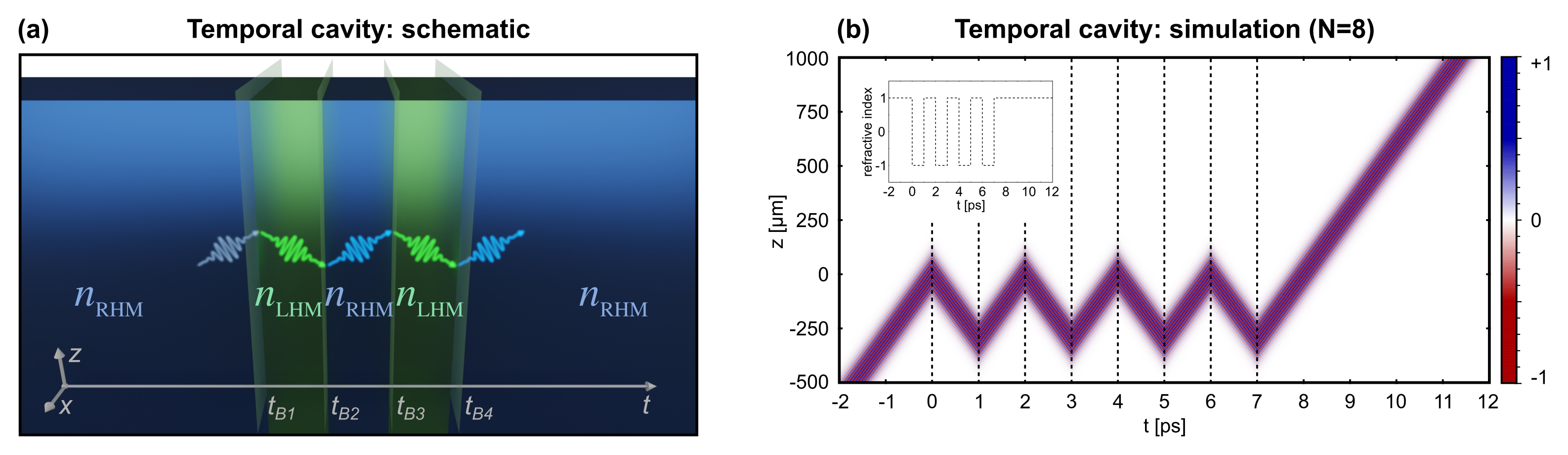}
	\caption{\textbf{Temporal cavity.} \textbf{(a)} Schematic representation of a temporal cavity based on a sequence of NF-LHTBs alternating RHM and NF-LHM, so that the light bounce back and forth along the time. \textbf{(b)} E-field intensity of a Gaussian light pulse propagating through a temporal cavity. As indicated in the inset, the simulation is for a sequence of $N=8$ NF-LHTBs separated from each other a period of $\Delta T=1$~ps.}
	\label{Fig.04}
\end{figure*}

\begin{figure*}[t!]
	\centering
	\includegraphics[width=0.925\linewidth]{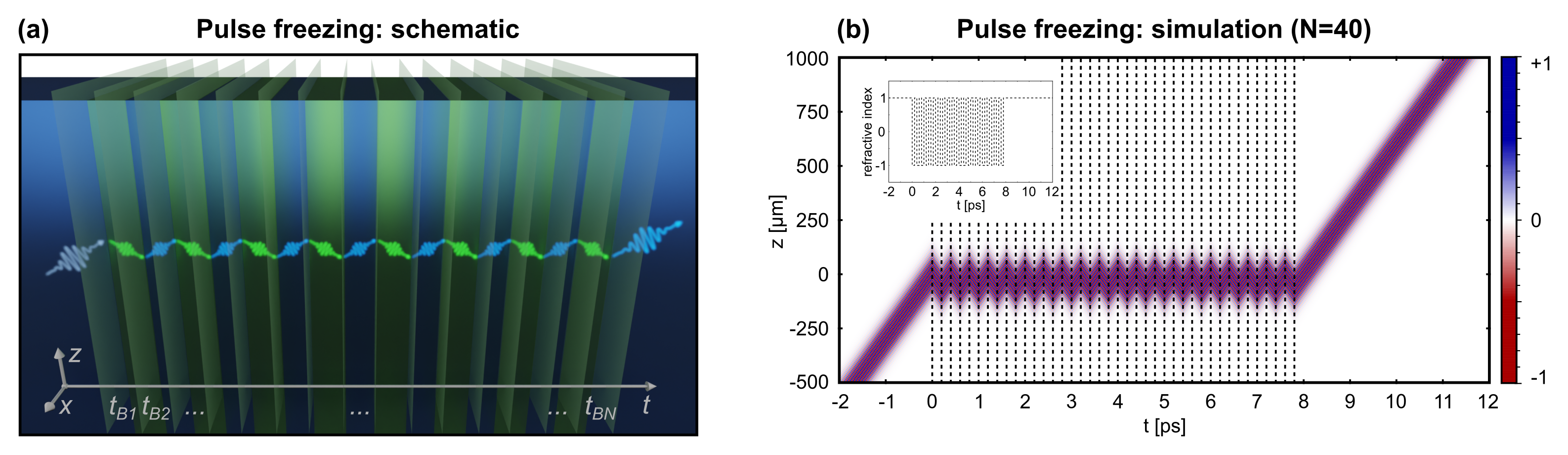}
	\caption{\textbf{Pulse freezing.} \textbf{(a)} Representation of a configuration based on a sequence of NF-LHTBs, alternating RHM and~NF-LHM, enabling pulse freezing. \textbf{(b)} E-field intensity of a Gaussian pulse spatially freezed within a sequence of $N=40$~NF-LHTBs separated from each other a period of $\Delta T=0.2$~ps.}
	\label{Fig.05}
\end{figure*}

Once conceptualized and mathematically modeled the temporal mirror above, the idea of a {\em temporal cavity} straightforwardly arises as a sequence of successive NF-LHTBs~[see~\hyperref[Fig.04]{Fig.~4(a)}]. That is, a set of temporal mirrors lined up and separated from each other a time interval such that each of them is induced after the EM wave has travelled a distance $\Delta z$. From the above insights, the overall effect of a temporal cavity will be that of a light pulse propagating and bouncing back after each temporal boundary. Importantly, just like for the temporal mirror, such an ideal behavior (disregarding spurious leaky radiation), relies on the impedance-matching condition at each time interface. Hence, according to the aforementioned labeling criteria for NF-LHTBs, the only signal that exists after each temporal boundary is that referred to as the forward wave having a reversed direction of propagation.

To numerically illustrate such an effect, we consider a scenario similar to the previous case for temporal mirrors, where a Gaussian pulse, whose electric component is linearly polarized on the $x$-axis, propagates in~vacuum~along the $z$-axis with $\omega=100/2\pi$~THz and $T=200$~fs. In this case, we use a sequence of $8$ NF-LHTBs, alternating~vacuum and NF-LHM~characterized by $\varepsilon_2=-1$, $\mu_2=-1$ ($n_2=-1$, $Z_2=1$), separated from each other a period of time $\Delta T=1$~ps, thereby conforming a sequence of $7+2$ sections (including the regions of the input and the output signals). Each transition brings in a temporal Fresnel reflection coefficient $r_{\rm TB}^{\rm NF}=-1$ modulating the field amplitude, and a variation in the phase that reverses the direction of propagation, thus switching the propagation between the positive and the negative direction along the $z$-axis. For this specific configuration, the electric field in each interval, i.e., in each temporal segment $(i-2)\Delta T\to (i-1)\Delta T$ with $i=\left\{2,N+1=9\right\}$ reads as:
\begin{equation}
{\bf E}_i(z,t)=\left[r_{\rm TB}^{\rm NF}\right]^i\left[\cos{(\varphi_i)}e^{-\left[\varphi_i^2/(2\omega^2T^2)\right]}\right]\hat{\bf x},
\end{equation}
where $\varphi_1=\varphi_{i-1}+\omega t[1-(i-2)\Delta T/t][v_{i-1}-v_i]/v_1$. Furthermore, it should be noted that, by definition, ${\bf E}_{\rm input}\equiv{\bf E}_1(z,t)$ and ${\bf E}_{\rm output}\equiv{\bf E}_9(z,t)$, and likewise $\varphi_{\rm input}\equiv\varphi_1$ and $\varphi_{\rm output}\equiv\varphi_9$, which are obtained from the above-mentioned rule of recurrence. By construction, in this particular case we have that:
\begin{equation}
v_i=\left\{\begin{matrix}
v_1=+c, & & \text{for }i\in\text{odd}, \\
v_2=-c, & & \text{for }i\in\text{even}, \\
\end{matrix}\right.
\end{equation}
so that, the phase can be recast as $\varphi_i=\varphi_{i-1}+\omega t[1-(i-2)\Delta T/t][(-1)^{i-1}\Delta n]$, i.e., only depending on~the temporal period $\Delta T$, and the index contrast $\Delta n=n_2-n_1$. From the above expressions, in~\hyperref[Fig.04]{Fig.~4(b)} we represent the evolution map of the electric field intensity for this particular sequence of $8$ NF-LHTBs conforming a temporal cavity. This juxtaposition of temporal mirrors can be conceived as a mechanism of temporal light confinement during the time that the temporally-modulated active region is switching the handedness of the medium. Finally, it is worth remarking that, akin to the temporal mirror, each NF-LHTB produces a $\pi$-shift in the phase that sharply reverses the field amplitude~\cite{Pacheco-Pena2023B}.

\subsection{Pulse freezing}
\label{Sect.IV.b}

\begin{figure*}[t!]
	\centering
	\includegraphics[width=0.95\linewidth]{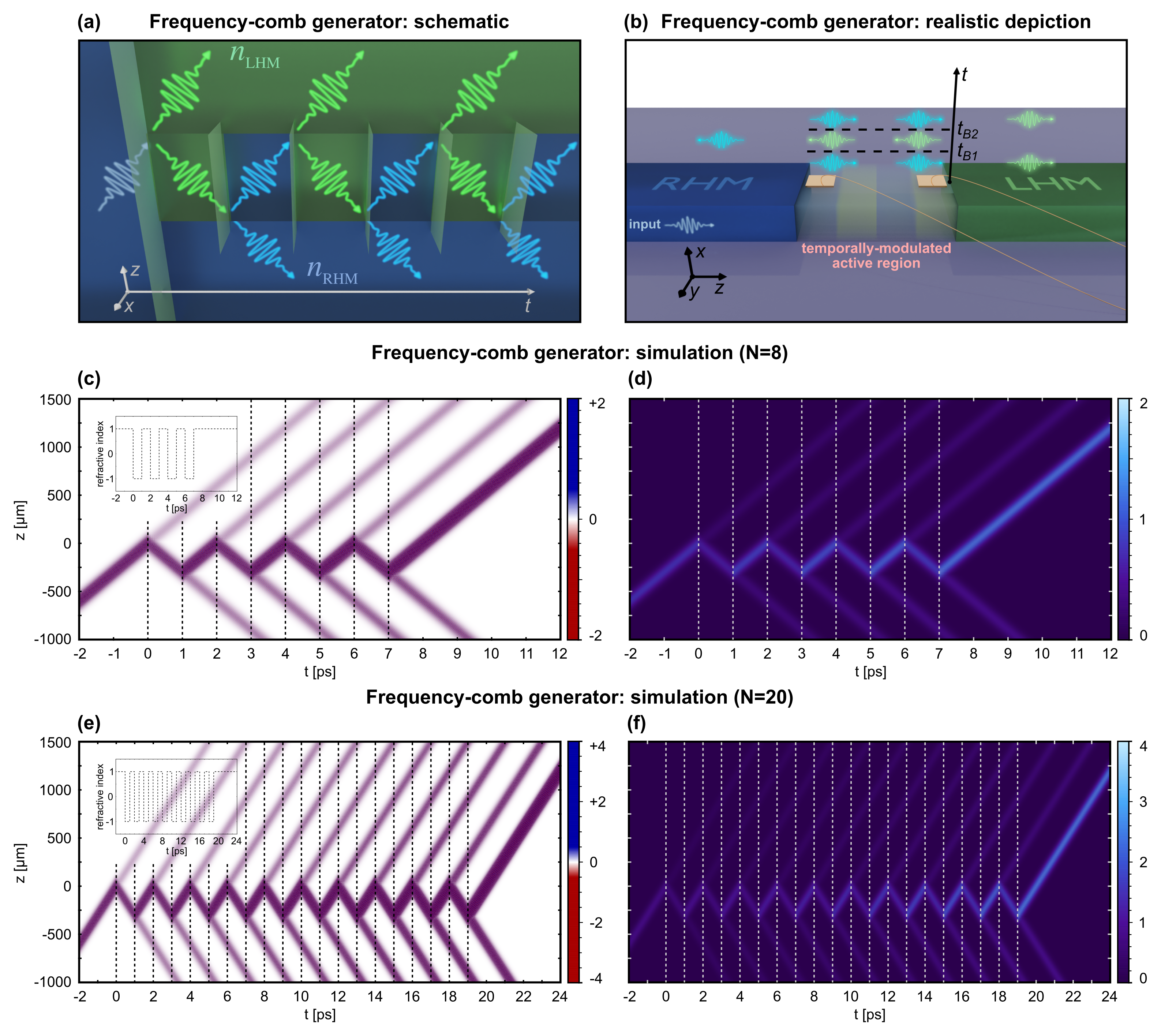}
	\caption{\textbf{Frequency-comb generator.} \textbf{(a)} Representation of a sophisticated design enabling a frequency-comb generator. After each temporal boundary, part of the radiation bounces back, while the remaining is released out the temporally active region, thereby freely propagating either through RHM (bottom medium, in blue) or NF-LHM (top medium, in green). \textbf{(b)}~Conceptual depiction of a realistic configuration to implement the frequency-comb generator, where the temporally-modulated active region, is placed in between a RHM and a NF-LHM. \textbf{(c)}~E-field intensity of a Gaussian light pulse for a sequence of $N=8$ NF-LHTBs separated from each other a period of $\Delta T=1$~ps, showing the effect of a frequency-comb generator, \textbf{(d)}~and the representation in absolute value for a better visualization of the energetic gain due to the temporal switching. \textbf{(e-f)}~Same representation for a sequence of $N=20$ NF-LHTBs.}
	\label{Fig.06}
\end{figure*}

As a stemming application of the temporal cavity, we put forward a configuration enabling {\em pulse freezing}~[see~\hyperref[Fig.05]{Fig.~5(a)}]. Such an effect would result from shrinking the internal period of the temporal cavity so that the pulses do not have enough time to bounce back and propagate forth between each temporal transition. Instead, they appear to be static (temporally frozen) in a spatial region while the temporally-modulated region is active.

Following the previous mathematical modeling for the input field, in~\hyperref[Fig.05]{Fig.~5(b)} we numerically represent the evolution of the electric field intensity for a particular configuration consisting in a sequence of $40$ NF-LHTBs, i.e., $39+2$ sections (including the regions where the input and the output signals propagate) alternating the vacuum and the same NF-LHM as above, separated from each other a period of time $\Delta T=0.2$~ps. We can observe that in each temporal interval, the EM wave does not have enough time to spatially travel, meaning that the light pulse is spatially static.

\subsection{Frequency-comb generator}
\label{Sect.IV.c}

Finally, upon the basis of the above configurations enabled by NF-LHTBs, we propose a {\em frequency-comb generator}, which can be conceived as the precursor of a {\em temporal laser}~[see~\hyperref[Fig.06]{Fig.~6(a)}]. This functionality essentially relies on the temporal cavity, but with a peculiarity. While part of the EM radiation bounces back after each temporal boundary, the remaining signal is released from the temporally active region. To reach this behavior, as schematically shown in~\hyperref[Fig.06]{Fig.~6(b)}, the temporally-modulated active region is to be constrained to a spatially bounded region, sandwiched between a RHM and a NF-LHM semi-infinite spatial regions. Likewise, in contrast with the previous functionalities which require the fulfillment of the impedance-matching condition, in this case, such a condition must be necessarily dismissed. This is to allow the signal to be released. Thus, once the signals leave the temporally-modulated active region, they freely propagate along a direction depending on whether the transition is either from RHM to NF-LHM, keeping the same direction, or from NF-LHM to RHM, reversing the propagation with respect to the input signal. Furthermore, a remarkable consequence of the impedance mismatch is that the temporal transitions bring about changes on the EM energy of the input signal~\cite{Pacheco-Pena2018,Pacheco-Pena2023B}, which shall decrease or increase, respectively, for RHM to NF-LHM or NF-LHM to RHM temporal transitions. This rate of energy variation shall depend on the specific characteristics of the medium, but, at any rate, once the signal is released out the active region, since we are assuming non-dispersive (and hence lossless/gainless) media, the EM energy will keep constant, and the EM wave will freely propagate throughout each output media.

To numerically illustrate this application, we consider a similar scenario as above, namely, a linearly polarized~Gaussian pulse propagating along the $z$-axis with $\omega=100/2\pi$~THz and $T=200$~fs. In this case, though, the impedance mismatch is set by means of a sequence of successive NF-LHTBs alternating a vacuum and a NF-LHM characterized by $\varepsilon_2=-2$, $\mu_2=-1/2$ ($n_2=-1$, $Z_2=1/2$), so that the temporal Fresnel reflection and transmission coefficients~are given~by $r_{\rm TB}^{\rm NF}=-3/4$ and $t_{\rm TB}^{\rm NF}=1/4$ for NF-LHTB between~RHM to NF-LHM, and $r_{\rm TB}^{\rm NF}=-3/2$ and $t_{\rm TB}^{\rm NF}=-1/2$ for NF-LHTBs between NF-LHM to RHM. Upon this general configuration, we present two particular examples, respectively consisting in a sequence of $8$~[see~\hyperref[Fig.06]{Fig.~6(c)}] and $20$~[see~\hyperref[Fig.06]{Fig.~6(e)}] NF-LHTBs, namely, $7+2$ and $19+2$ sections, including the regions of the input and the output signals, separated from each other a period of time $\Delta T=1$~ps. In both cases it can be clearly seen that, after each temporal boundary, the EM pulse is split into two parts: one signal that bounces back standing in the temporally-modulated active region, and another component that keeps its propagation along the same direction prior to the temporal boundary. Notice that, in this particular case, since the refractive index is set with the same modulus both in the RHM and the NF-LHM, $|n_1|=|n_2|=1$, the velocity of propagation remains constant. Finally, it is also worth to highlight the variation of the EM energy after each NF-LHTB, which becomes more clearly visualized by means of the absolute value representation~[see~\hyperref[Fig.06]{Figs.~6(d,f)}].

\section{Conclusions}
\label{Sect.V}

In this work, we have addressed the fundamental question on whether a mirror without temporal boundaries (i.e., an unbounded platform able to produce total reflection without the need of spatial boundaries) can exist. Based on the conservation of the Minkowski momentum, we have theoretically demonstrated that perfect temporal mirrors are in general forbidden, but they might be physically feasible for a special class of artificially engineered metamaterials. Specifically, we have showed that active materials whose Minkowski momentum and energy flow (determined by the group velocity) are anti-parallel, can be used to implement purely temporal mirrors. We identified that non-Foster left-handed media satisfy such a condition, thereby enabling the realization of mirrors without spatial boundaries. Besides temporal mirrors, we have also shown that this class of non-Foster left-handed temporal boundaries allows for the possibility to perform other related photonic applications, including temporal cavities, pulse freezing, and frequency-comb generators. Beyond expanding the portfolio of time-varying photonic functionalities, this study offers insights into fundamental physical aspects, such as the very nature of a mirror.

\section{Appendix A: Derivation of the Fresnel coefficients for conventional temporal boundaries}
\label{AppendixA}

The Fresnel coefficients associated to conventional right-handed temporal boundaries, can be straightforwardly obtained just by properly impossing the continuity of the displacement vector field, ${\bf D}(z,t)$, and the magnetic induction field, ${\bf B}(z,t)$, at the temporal interface~\cite{Morgenthaler1958,Fante1958,Xiao2014,Mendonca2002,Zurita-Sanchez2009,Plansinis2015,Ramaccia2020,Gratus2021,Mai2023,Mostafa2024}, namely, ${\bf D}(z,t_{\rm B}^{-})={\bf D}(z,t_{\rm B}^{+})$ and ${\bf B}(z,t_{\rm B}^{-})={\bf B}(z,t_{\rm B}^{+})$, where the temporal boundary is assumed to be at $t=t_{\rm B}$. From this, we consider the following input fields:
\begin{align}
&{\bf E}_{\rm input}(z,t)=E_0e^{-i\omega_1 t}e^{+ik_1 z}\hat{\bf x},\\
&{\bf H}_{\rm input}(z,t)=\frac{E_0}{Z_1}e^{-i\omega_1 t}e^{+ik_1 z}\hat{\bf y}.
\end{align}
After the temporal boundary, which, for simplicity and without any loss of generality, it is assumed to occur at $t=t_{\rm B}=0$, both the permittivity and the permeability, and hence the refractive index and the impedance, undergo a rapid change, so that $\varepsilon_1\to\varepsilon_2$, $\mu_1\to\mu_2$, and $n_1\to n_2$, $Z_1\to Z_2$. Notice that, in this case, all the parameters are supposed to be real and positive. Hence, the output fields can be generally expressed as:
\begin{align}
&\!\!{\bf E}_{\rm output}(z,t)\!=\!E_0\left[t_{\rm TB}e^{+i\omega_2 t}+r_{\rm TB}e^{-i\omega_2 t}\right]e^{+ik_2z}\hat{\bf x},\\
&\!\!{\bf H}_{\rm output}(z,t)\!=\!\frac{E_0}{Z_2}\left[t_{\rm TB}e^{+i\omega_2 t}-r_{\rm TB}e^{-i\omega_2 t}\right]e^{+ik_2z}\hat{\bf y}.
\end{align}
Imposing the continuity of ${\bf D}=\varepsilon{\bf E}$ and ${\bf B}=\mu{\bf H}$ at the temporal interface, it follows that:
\begin{align}
\!\!\!\!{\bf D}_{\rm input}(z,0)\!=\!{\bf D}_{\rm output}(z,0)&\rightarrow\varepsilon_1\!=\!\varepsilon_2\!\left[t_{\rm TB}+r_{\rm TB}\right]\!,\!\!\\
\!\!\!\!{\bf B}_{\rm input}(z,0)\!=\!{\bf B}_{\rm output}(z,0)&\rightarrow\frac{\mu_1}{Z_1}\!=\!\frac{\mu_2}{Z_2}\!\left[t_{\rm TB}-r_{\rm TB}\right]\!.\!\!
\end{align}
Thus, the corresponding temporal Fresnel reflection and transmission coefficients for conventional right-handed temporal boundaries are given by:
\begin{align}
r_{\rm TB}=\frac{1}{2}\left[\frac{\varepsilon_1}{\varepsilon_2}-\frac{n_1}{n_2}\right]=\frac{\varepsilon_1}{\varepsilon_2}\left[\frac{Z_2-Z_1}{2Z_2}\right],\\
t_{\rm TB}=\frac{1}{2}\left[\frac{\varepsilon_1}{\varepsilon_2}+\frac{n_1}{n_2}\right]=\frac{\varepsilon_1}{\varepsilon_2}\left[\frac{Z_2+Z_1}{2Z_2}\right].
\end{align}

\section{Appendix B: Derivation of the Fresnel coefficients for NF-LHTBs}
\label{AppendixB}

To derive the Fresnel coefficients for a NF-LHTB we depart from the same input fields:
\begin{align}
&{\bf E}_{\rm input}(z,t)=E_0e^{-i\omega_1 t}e^{+ik_1 z}\hat{\bf x},\\
&{\bf H}_{\rm input}(z,t)=\frac{E_0}{Z_1}e^{-i\omega_1 t}e^{+ik_1 z}\hat{\bf y}.
\end{align}
Again, it is assumed that the temporal transition occurs at $t=t_{\rm B}=0$, so that the permittivity and the permeability, and hence the refractive index and the impedance, undergo a rapid change as indicated above, that is $\varepsilon_1\to\varepsilon_2$, $\mu_1\to\mu_2$, and $n_1\to n_2$, $Z_1\to Z_2$. In this case, since we are considering a NF-LHTB, the constitutive parameters after the temporal boundary, that is, $\varepsilon_2$ and $\mu_2$, and consequently the refractive index, $n_2$ (but not so the impedance, $Z_2$), are assumed to be real and negative. Hence, the output fields read as:
\begin{align}
&\!\!{\bf E}_{\rm output}(z,t)\!=\!E_0\left[t_{\rm TB}^{\rm NF}e^{-i\omega_2 t}+r_{\rm TB}^{\rm NF}e^{+i\omega_2 t}\right]e^{+ik_2z}\hat{\bf x},\\
&\!\!{\bf H}_{\rm output}(z,t)\!=\!\frac{E_0}{Z_2}\left[r_{\rm TB}^{\rm NF}e^{+i\omega_2 t}-t_{\rm TB}^{\rm NF}e^{-i\omega_2 t}\right]e^{+ik_2z}\hat{\bf y}.
\end{align}
The continuity of ${\bf D}$ and ${\bf B}$ at the temporal interface leads to:
\begin{align}
\!\!\!\!{\bf D}_{\rm input}(z,0)\!=\!{\bf D}_{\rm output}(z,0)&\rightarrow\varepsilon_1\!=\!\varepsilon_2\!\left[t_{\rm TB}^{\rm NF}+r_{\rm TB}^{\rm NF}\right]\!,\!\!\\
\!\!\!\!{\bf B}_{\rm input}(z,0)\!=\!{\bf B}_{\rm output}(z,0)&\rightarrow\frac{\mu_1}{Z_1}\!=\!\frac{\mu_2}{Z_2}\!\left[r_{\rm TB}^{\rm NF}-t_{\rm TB}^{\rm NF}\right]\!.\!\!
\end{align}
Thus, the reflection and transmission Fresnel coefficients for NF-LHTBs are:
\begin{align}
r_{\rm TB}^{\rm NF}=\frac{1}{2}\left[\frac{\varepsilon_1}{\varepsilon_2}+\frac{n_1}{n_2}\right]=\frac{\varepsilon_1}{\varepsilon_2}\left[\frac{Z_2+Z_1}{2Z_2}\right],\\
t_{\rm TB}^{\rm NF}=\frac{1}{2}\left[\frac{\varepsilon_1}{\varepsilon_2}-\frac{n_1}{n_2}\right]=\frac{\varepsilon_1}{\varepsilon_2}\left[\frac{Z_2-Z_1}{2Z_2}\right].
\end{align}
As we can observe, the reflection and transmission Fresnel coefficients for NF-LHTBs are exactly swapped with respect to the case of conventional right-handed temporal boundaries.
\vspace{-0.05cm}


\section{Acknowledgments}
The authors would like to thank Prof. Nader Engheta for his fundamental insights on the original conception and fruitful discussions on the idea of temporal mirrors without spatial boundaries.
This work was supported by ERC Starting Grant No.~ERC-2020-STG-948504-NZINATECH.
J.E.V.-L. acknowledges support from Juan de la Cierva--Formaci\'on fellowship~FJC2021-047776-I and project PJUPNA2025-11905.
V.P.-P. acknowledges support from the Leverhulme Trust under the Leverhulme Trust Research Project Grant Scheme (No. RPG-2020-316, No. RPG-2023-024).
I.L. further acknowledges support from Ram\'on y Cajal fellowship~RYC2018-024123-I and project RTI2018-093714-301J-I00 (MCIU/AEI/FEDER/UE).



{\small
\begin{thebibliography}{90}
\setlength{\itemsep}{1ex}
\label{Sect.Ref}

\bibitem{Yin2022} Yin, S., Galiffi, E., \& Al\`u, A. Floquet metamaterials. \href{https://doi.org/10.1186/s43593-022-00015-1}{\textit{eLight} \textbf{2}, 1 (2022)}.

\bibitem{Galiffi2022} Galiffi, E., Tirole, R., Yin, S., Li, H., Vezzoli, S., Huidobro, P. A., Silveirinha, M. G., Sapienza, R., Al\`u, A., \& Pendry, J. Photonics of time-varying media. \href{https://doi.org/10.1117/1.AP.4.1.014002}{\textit{Adv. Photonics} \textbf{4}, 014002 (2022)}.

\bibitem{Caloz2020A} Caloz, C. \& Deck-L\'eger, Z.-L. Spacetime metamaterials—part I: General concepts. \href{https://doi.org/10.1109/TAP.2019.2944225}{\textit{IEEE Trans. Antennas Propag.} \textbf{68}, 1569 (2020)}.

\bibitem{Caloz2020B} Caloz, C. \& Deck-L\'eger, Z.-L. Spacetime metamaterials—part II: Theory and applications. \href{https://doi.org/10.1109/TAP.2019.2944216}{\textit{IEEE Trans. Antennas Propag.} \textbf{68}, 1583 (2020)}.

\bibitem{Caloz2022} Caloz, C., Deck-L\'eger, Z.-L., Bahrami, A., C\'espedes, O., \& Li, Z. Generalized space-time engineered modulation (GSTEM) metamaterials: A global and extended perspective. \href{https://doi.org/10.1109/MAP.2022.3216773}{\textit{IEEE Antennas Propag. Mag.} \textbf{65}, 50 (2023)}.

\bibitem{Engheta2021} Engheta, N. Metamaterials with high degrees of freedom: space, time, and more. \href{https://doi.org/10.1515/nanoph-2020-0414}{\textit{Nanophotonics} \textbf{10}, 639 (2021)}.

\bibitem{Yuan2022} Yuan, L. \& Fan, S. Temporal modulation brings metamaterials into new era. \href{https://doi.org/10.1038/s41377-022-00870-0}{\textit{Light Sci. Appl.} \textbf{11}, 173 (2022)}.

\bibitem{Engheta2023} Engheta, N. Four-dimensional optics using time-varying metamaterials. \href{https://doi.org/10.1126/science.adf1094}{\textit{Science} \textbf{379}, 1190 (2023)}.

\bibitem{Taravati2020} Taravati, S. \& Kishk, A. A. Space-time modulation: Principles and applications. \href{https://doi.org/10.1109/MMM.2019.2963606}{\textit{IEEE Microwave} \textbf{21}, 30 (2020)}.

\bibitem{Shaltout2019A} Shaltout, A. M., Shalaev, V. M., \& Brongersma, M. L. Spatiotemporal light control with active metasurfaces. \href{https://doi.org/10.1126/science.aat3100}{\textit{Science} \textbf{364}, 648 (2019)}.

\bibitem{Pacheco-Pena2023A} Pacheco-Pe\~na, V., Fink, M., \& Engheta, N. Temporal chirp, temporal lensing and temporal routing via space-time interfaces. \href{https://arxiv.org/abs/2311.10855}{arXiv:2311.10855 (2023)}.

\bibitem{Pacheco-Pena2024} Pacheco-Pe\~na, V. \& Engheta, N. Spatiotemporal cascading of dielectric waveguides [Invited]. \href{https://doi.org/10.1364/OME.516262}{\textit{Opt. Mater. Express} \textbf{14}, 1062 (2024)}.

\bibitem{Yoon2016} Yoon, G., Kim, I., \& Rho, J. Challenges in fabrication towards realization of practical metamaterials. \href{https://doi.org/10.1016/j.mee.2016.05.005}{\textit{Microelectron. Eng.} \textbf{163}, 7 (2016)}.

\bibitem{Hayran2023} Hayran, Z. \& Monticone, F. Using time-varying systems to challenge fundamental limitations in electromagnetics: Overview and summary of applications. \href{https://doi.org/10.1109/MAP.2023.3236275}{\textit{IEEE Antennas Propag. Mag.} \textbf{65}, 29 (2023)}.

\bibitem{Pacheco-Pena2022} Pacheco-Pe\~na, V., Sol\'is, D. M., \& Engheta, N. Time-varying electromagnetic media: opinion. \href{https://doi.org/10.1364/OME.471007}{\textit{Opt. Mater. Express} \textbf{12}, 3829 (2022)}.

\bibitem{Sounas2017} Sounas, D. L. \& Al\`u, A. Non-reciprocal photonics based on time modulation. \href{https://doi.org/10.1038/s41566-017-0051-x}{\textit{Nat. Photon.} \textbf{11}, 774 (2017)}.

\bibitem{Li2022} Li, H., Yin, S., \& Al\`u, A. Nonreciprocity and Faraday rotation at time interfaces. \href{https://doi.org/10.1103/PhysRevLett.128.173901}{\textit{Phys. Rev. Lett.} \textbf{128}, 173901 (2022)}.

\bibitem{Solis2021A} Sol\'is, D. M. \& Engheta, N. Functional analysis of the polarization response in linear time-varying media: A generalization of the Kramers-Kronig relations. \href{https://doi.org/10.1103/PhysRevB.103.144303}{\textit{Phys. Rev. B} \textbf{103}, 144303 (2021)}.

\bibitem{Koutserimpas2024} Koutserimpas, T. T. \& Monticone, F. Time-varying media, dispersion, and the principle of causality [Invited]. \href{https://doi.org/10.1364/OME.515957}{\textit{Opt. Mater. Express} \textbf{14}, 1222 (2024)}.

\bibitem{Hayran2022} Hayran, Z., Khurgin, J. B., \& Monticone, F. $\hbar\omega$ versus $\hbar k$: dispersion and energy constraints on time-varying photonic materials and time crystals [Invited]. \href{https://doi.org/10.1364/OME.471672}{\textit{Opt. Mater. Express} \textbf{12}, 3904 (2022)}.

\bibitem{Pendry2021} Pendry, J. B., Galiffi, E., \& Huidobro, P. A. Gain in time-dependent media -- A new mechanism. \href{https://doi.org/10.1364/JOSAB.427682}{\textit{J. Opt. Soc. Am. B} \textbf{38}, 3360 (2021)}.

\bibitem{Bacot2016} Bacot, V., Labousse, M., Eddi, A., Fink, M., \& Fort, E. Time reversal and holography with spacetime transformations. \href{https://doi.org/10.1038/nphys3810}{\textit{Nat. Phys.} \textbf{12}, 972 (2016)}.

\bibitem{Dong2024} Dong, Z., Li, H., Wan, T., Liang, Q., Yang, Z., \& Yan, B. Quantum time reflection and refraction of ultracold atoms. \href{https://doi.org/10.1038/s41566-023-01290-1}{\textit{Nat. Photon.} \textbf{18}, 68 (2024)}.

\bibitem{Hadad2015} Hadad, Y., Sounas, D. L., \& Al\`u, A. Space-time gradient metasurfaces. \href{https://doi.org/10.1103/PhysRevB.92.100304}{\textit{Phys. Rev. B} \textbf{92}, 100304 (2015)}.

\bibitem{Martinez-Romero2016} Mart\'inez-Romero, J. S., Becerra-Fuentes, O. M., \& Halevi, P. Temporal photonic crystals with modulations of both permittivity and permeability. \href{https://doi.org/10.1103/PhysRevA.93.063813}{\textit{Phys. Rev. A} \textbf{93}, 063813 (2016)}.

\bibitem{Shaltout2019B} Shaltout, A. M., Shalaev, V. M., \& Brongersma, M. L. Spatiotemporal light control with active metasurfaces. \href{https://doi.org/10.1126/science.aat3100}{\textit{Science} \textbf{364}, eaat3100 (2019)}.

\bibitem{Pacheco-Pena2020A} Pacheco-Pe\~na, V. \& Engheta, N. Effective-medium concepts in temporal metamaterials. \href{https://doi.org/10.1515/nanoph-2019-0305}{\textit{Nanophotonics} \textbf{9}, 379 (2020)}.

\bibitem{Huidobro2021} Huidobro, P. A., Silveirinha, M. G., Galiffi, E., \& Pendry, J. B. Homogenization theory of space-time metamaterials. \href{https://doi.org/10.1103/PhysRevApplied.16.014044}{\textit{Phys. Rev. Appl.} \textbf{16}, 014044 (2021)}.

\bibitem{Alex-Amor2023} Alex-Amor, A., Molero, C., \& Silveirinha, M. G. Analysis of metallic space-time gratings using Lorentz transformations. \href{https://doi.org/10.1103/PhysRevApplied.20.014063}{\textit{Phys. Rev. Appl.} \textbf{20}, 014063 (2023)}.

\bibitem{Akbarzadeh2018} Akbarzadeh, A., Chamanara, N., \& Caloz, C. Inverse prism based on temporal discontinuity and spatial dispersion. \href{https://doi.org/10.1364/OL.43.003297}{\textit{Opt. Lett.} \textbf{43}, 3297 (2018)}.

\bibitem{Pacheco-Pena2020B} Pacheco-Pe\~na, V. \& Engheta, N. Temporal aiming. \href{https://doi.org/10.1038/s41377-020-00360-1}{\textit{Light Sci. Appl.} \textbf{9}, 1 (2020)}.

\bibitem{Pacheco-Pena2020C} Pacheco-Pe\~na, V. \& Engheta, N. Antireflection temporal coatings. \href{https://doi.org/10.1364/OPTICA.381175}{\textit{Optica} \textbf{7}, 323 (2020)}.

\bibitem{Ramaccia2021} Ramaccia, D., Al\`u, A., Toscano, A., \& Bilotti, F. Temporal multilayer structures for designing higher-order transfer functions using time-varying metamaterials. \href{https://doi.org/10.1063/5.0042567}{\textit{Appl. Phys. Lett.} \textbf{118}, 101901 (2021)}.

\bibitem{Castaldi2021} Castaldi, G., Pacheco-Pe\~na, V., Moccia, M., Engheta, N., \& Galdi, V. Exploiting space-time duality in the synthesis of impedance transformers via temporal metamaterials. \href{https://doi.org/10.1515/nanoph-2021-0231}{\textit{Nanophotonics} \textbf{10}, 3687 (2021)}.

\bibitem{Pacheco-Pena2021} Pacheco-Pe\~na, V. \& Engheta, N. Temporal equivalent of the Brewster angle. \href{https://doi.org/10.1103/PhysRevB.104.214308}{\textit{Phys. Rev. B} \textbf{104}, 214308 (2021)}.

\bibitem{Wang2023} Wang, X., Mirmoosa, M. S., Asadchy, V. S., Rockstuhl, C., Fan, S., \& Tretyakov, S. A. Metasurface-based realization of photonic time crystals. \href{https://doi.org/10.1126/sciadv.adg7541}{\textit{Sci. Adv.} \textbf{9}, eadg7541 (2023)}.

\bibitem{Mirmoosa2019} Mirmoosa, M. S., Ptitcyn, G., Asadchy, V. S., \& Tretyakov, S. A. Time-varying reactive elements for extreme accumulation of electromagnetic energy. \href{https://doi.org/10.1103/PhysRevApplied.11.014024}{\textit{Phys. Rev. Appl.} \textbf{11}, 014024 (2019)}.

\bibitem{Vazquez-Lozano2023A} V\'azquez-Lozano, J. E. \& Liberal, I. Incandescent temporal metamaterials. \href{https://doi.org/10.1038/s41467-023-40281-2}{\textit{Nat. Commun.} \textbf{14}, 4606 (2023)}.

\bibitem{Mendonca2000} Mendon\c{c}a, J. T., Guerreiro, A., \& Martins, A. M. Quantum theory of time refraction. \href{https://doi.org/10.1103/PhysRevA.62.033805}{\textit{Phys. Rev. A} \textbf{62}, 033805 (2000)}.

\bibitem{Mendonca2005} Mendon\c{c}a, J. T., \& Guerreiro, A. Time refraction and the quantum properties of vacuum. \href{https://doi.org/10.1103/PhysRevA.72.063805}{\textit{Phys. Rev. A} \textbf{72}, 063805 (2005)}.

\bibitem{Kort-Kamp2021} Kort-Kamp, W. J. M., Azad, A. K., \& Dalvit, D. A. R. Space-time quantum metasurfaces. \href{https://doi.org/10.1103/PhysRevLett.127.043603}{\textit{Phys. Rev. Lett.} \textbf{127}, 043603 (2021)}.

\bibitem{Vazquez-Lozano2023B} V\'azquez-Lozano, J. E. \& Liberal, I. Shaping the quantum vacuum with anisotropic temporal boundaries. \href{https://doi.org/10.1515/nanoph-2022-0491}{\textit{Nanophotonics} \textbf{12}, 539 (2023)}.

\bibitem{Liberal2023} Liberal, I., V\'azquez-Lozano, J. E., \& Pacheco-Pe\~na, V. Quantum antireflection temporal coatings: quantum state frequency shifting and inhibited thermal noise amplification. \href{https://doi.org/10.1002/lpor.202200720}{\textit{Laser Photon. Rev.} \textbf{17}, 2200720 (2023)}.

\bibitem{Tirole2022} Tirole, R., Galiffi, E., Dranczewski, J., Attavar, T., Tilmann, B., Wang, Y.-T., Huidobro, P. A., Al\`u, A., Pendry, J. B., Maier, S. A., Vezzoli, S., \& Sapienza, R. Saturable time-varying mirror based on an epsilon-near-zero material. \href{https://doi.org/10.1103/PhysRevApplied.18.054067}{\textit{Phys. Rev. Appl.} \textbf{18}, 054067 (2022)}.

\bibitem{Tirole2023} Tirole, R., Vezzoli, S., Galiffi, E., Robertson, I., Maurice, D., Tilmann, B., Maier, S. A., Pendry, J. B., \& Sapienza, R. Double-slit time diffraction at optical frequencies. \href{https://doi.org/10.1038/s41567-023-01993-w}{\textit{Nat. Phys.} \textbf{19}, 999 (2023)}.

\bibitem{Moussa2023} Moussa, H., Xu, G., Yin, S., Galiffi, E., Ra'di, Y., \& Al\`u, A. Observation of temporal reflection and broadband frequency translation at photonic time interfaces. \href{https://doi.org/10.1038/s41567-023-01975-y}{\textit{Nat. Phys.} \textbf{19}, 863 (2023)}.

\bibitem{Nicholls2017} Nicholls, L. H., Rodr\'iguez-Fortu\~no, F. J., Nasir, M. E., C\'orova-Castro, R. M., Olivier, N., Wurtz, G. A., \& Zayats, A. V. Ultrafast synthesis and switching of light polarization in nonlinear anisotropic metamaterials. \href{https://doi.org/10.1038/s41566-017-0002-6}{\textit{Nat. Photon.} \textbf{11}, 628 (2017)}.

\bibitem{Born} Born, M. \& Wolf, E. \textit{Principles of Optics} (Pergamon, 2005).

\bibitem{Novotny} Novotny, L. \& Hecht, B. \textit{Principles of Nano-Optics} (Cambridge University Press, 2012).

\bibitem{Morgenthaler1958} Morgenthaler, F. R. Velocity modulation of electromagnetic waves. \href{https://doi.org/10.1109/TMTT.1958.1124533}{\textit{IEEE Trans. Microwave Theory Tech.} \textbf{6}, 167 (1958)}.

\bibitem{Fante1958} Fante, R. L. Transmission of electromagnetic waves into time-varying media. \href{https://doi.org/10.1109/TAP.1971.1139931}{\textit{IEEE Trans. Antennas Propag.} \textbf{19}, 417 (1958)}.

\bibitem{Xiao2014} Xiao, Y., Maywar, D. N., \& Agrawal, G. P. Reflection and transmission of electromagnetic waves at a temporal boundary. \href{https://doi.org/10.1364/OL.39.000574}{\textit{Opt. Lett.} \textbf{39}, 574 (2014)}.

\bibitem{Mendonca2002} Mendon\c{c}a, J. T. \& Shukla, P. K. Time refraction and time reflection: Two basic concepts. \href{https://doi.org/10.1238/Physica.Regular.065a00160}{\textit{Phys. Scr.} \textbf{65}, 160 (2002)}.

\bibitem{Zurita-Sanchez2009} Zurita-S\'anchez, J. R., Halevi, P., \& Cervantes-Gonz\'alez, J. C. Reflection and transmission of a wave incident on a slab with a time-periodic dielectric function $\epsilon(t)$. \href{https://doi.org/10.1103/PhysRevA.79.053821}{\textit{Phys. Rev. A} \textbf{79}, 053821 (2009)}.

\bibitem{Plansinis2015} Plansinis, B. W., Donaldson, W. R., \& Agrawal, G. P. What is the temporal analog of reflection and refraction of optical beams? \href{https://doi.org/10.1103/PhysRevLett.115.183901}{\textit{Phys. Rev. Lett.} \textbf{115}, 183901 (2015)}.

\bibitem{Ramaccia2020} Ramaccia, D., Toscano, A., \& Bilotti, F. Light propagation through metamaterial temporal slabs: reflection, refraction, and special cases. \href{https://doi.org/10.1364/OL.402856}{\textit{Opt. Lett.} \textbf{45}, 5836 (2020)}.

\bibitem{Gratus2021} Gratus, J., Seviour, R., Kinsler, P., \& Jaroszynski, D. A. Temporal boundaries in electromagnetic materials. \href{https://doi.org/10.1088/1367-2630/ac1896}{\textit{New J. Phys.} \textbf{23}, 083032 (2021)}.

\bibitem{Mai2023} Mai, W., Xu, J., \& Werner, D. H. Fundamental asymmetries between spatial and temporal boundaries in electromagnetics. \href{https://doi.org/10.3390/sym15040858}{\textit{Symmetry} \textbf{15}, 858 (2023)}.

\bibitem{Mostafa2024} Mostafa, M. H., Mirmoosa, M. S., Sidorenko, M. S., Asadchy, V. S., \& Tretyakov, S. A. Temporal interfaces in complex electromagnetic materials: an overview [Invited]. \href{https://doi.org/10.1364/OME.516179}{\textit{Opt. Mater. Express} \textbf{14}, 1103 (2024)}.

\bibitem{Ptitcyn2023A} Ptitcyn, G., Mirmoosa, M. S., Sotoodehfar, A., \& Tretyakov, S. A. A tutorial on the basics of time-varying electromagnetic systems and circuits: Historic overview and basic concepts of time-modulation. \href{https://doi.org/10.1109/MAP.2023.3261601}{\textit{IEEE Antennas Propag. Mag.} \textbf{65}, 10 (2023)}.

\bibitem{Ortega-Gomez2023} Ortega-Gomez, A., Lobet, M., V\'azquez-Lozano, J. E., \& Liberal, I. Tutorial on the conservation of momentum in photonic time-varying media [Invited]. \href{https://doi.org/10.1364/OME.485540}{\textit{Opt. Mater. Express} \textbf{13}, 1598 (2023)}.

\bibitem{Hrabar2020} Hrabar, S. Time-varying route to non-Foster elements, in \href{https://doi.org/10.1109/Metamaterials49557.2020.9285065}{\textit{14th International Congress on Artificial Materials for Novel Wave Phenomena (Metamaterials)} (IEEE, 2020) pp. 108-109}.

\bibitem{Hrabar2022} Hrabar, S. Time-varying versus non-Foster elements - Similarities and differences, in \href{https://doi.org/10.1109/Metamaterials54993.2022.9920819}{\textit{16th International Congress on Artificial Materials for Novel Wave Phenomena (Metamaterials)} (IEEE, 2022) pp. 199-201}.

\bibitem{Ptitcyn2022} Ptitcyn, G., Mirmoosa, M. S., Hrabar, S., \& Tretyakov, S. A. Time-varying elements for realization of stable non-Foster circuits and metasurfaces, in \href{https://doi.org/10.1109/Metamaterials54993.2022.9920851}{\textit{16th International Congress on Artificial Materials for Novel Wave Phenomena (Metamaterials)} (IEEE, 2022) pp. 347-349}.

\bibitem{Kiasat2018} Kiasat, Y., Pacheco-Pe\~na, V., Edwards, B., \& Engheta, N. Temporal metamaterials with non-Foster networks, in \href{10.1364/CLEO_AT.2018.JW2A.90}{\textit{CLEO: Science and Innovations} (Optical Society of America, 2018) pp. JW2A-90}.

\bibitem{Pacheco-Pena2018} Pacheco-Pe\~na, V., Kiasat, Y., Edwards, B., \& Engheta, N. Salient features of temporal and spatio-temporal metamaterials, in \href{https://doi.org/10.1109/ICEAA.2018.8520356}{\textit{2018 International Conference on Electromagnetics and Advanced Applications (ICEAA)} (IEEE, 2018) pp. 524-526}.

\bibitem{Pacheco-Pena2023B} Pacheco-Pe\~na, V., Kiasat, Y., Sol\'is, D. M., Edwards, B., \& Engheta, N. Holding and amplifying electromagnetic waves with temporal non-Foster metastructures. \href{https://arxiv.org/abs/2304.03861}{arXiv:2304.03861 (2023)}.

\bibitem{Ptitcyn2023B} Ptitcyn, G. A., Mirmoosa, M. S., Hrabar, S., \& Tretyakov, S. A. Time-modulated circuits and metasurfaces for emulating arbitrary transfer functions. \href{https://doi.org/10.1103/PhysRevApplied.20.014041}{\textit{Phys. Rev. Appl.} \textbf{20}, 014041 (2023)}.

\bibitem{Lasri2023} Lasri, O. \& Sirota, L. Temporal negative refraction [Invited]. \href{https://doi.org/10.1364/OME.485242}{\textit{Opt. Mater. Express} \textbf{13}, 1401 (2023)}.

\bibitem{Tretyakov2001} Tretyakov, S. A. Meta-materials with wideband negative permittivity and permeability. \href{https://doi.org/10.1002/mop.1387}{\textit{Microw. Opt. Technol. Lett.} \textbf{31}, 163 (2001)}.

\bibitem{Engheta2005} Engheta, N. A positive future for double-negative metamaterials. \href{https://doi.org/10.1109/TMTT.2005.845188}{\textit{IEEE Trans. Microwave Theory Tech.} \textbf{53}, 1535 (2005)}.

\bibitem{Tretyakov2007} Tretyakov, S. A. \& Maslovski, S. I. Veselago Materials: What is possible and impossible about the dispersion of the constitutive parameters. \href{https://doi.org/10.1109/MAP.2007.370980}{\textit{IEEE Antennas Propag. Mag.} \textbf{49}, 37 (2007)}.

\bibitem{Solis2021B} Sol\'is, D. M., Kastner, R., \& Engheta, N. Time-varying materials in the presence of dispersion: plane-wave propagation in a Lorentzian medium with temporal discontinuity. \href{https://doi.org/10.1364/PRJ.427368}{\textit{Photonics Res.} \textbf{9}, 1842 (2021)}.

\bibitem{Zhang2021} Zhang, J., Donaldson, W. R., \& Agrawal, G. P. Temporal reflection and refraction of optical pulses inside a dispersive medium: an analytic approach. \href{https://doi.org/10.1364/JOSAB.416058}{\textit{J. Opt. Soc. Am. B} \textbf{38}, 997 (2021)}.

\bibitem{Hrabar2010A} Hrabar, S., Krois, I., Bonic, I., \& Kiricenko, A. Basic concepts of active dispersionless metamaterial based on non-Foster elements, in \href{https://ieeexplore.ieee.org/document/5729685}{\textit{20th International Conference on Applied Electromagnetics and Communications (ICECom)} (IEEE, 2010) pp. 1-4}.

\bibitem{Alvarez2020} Alvarez, J., Djafari-Rouhani, B., \& Torrent, D. Generalized elastodynamic model for nanophotonics. \href{https://doi.org/10.1103/PhysRevB.102.115308}{\textit{Phys. Rev. B} \textbf{102}, 115308 (2020)}.

\bibitem{Sakurai} Sakurai, J. J. \& Napolitano, J. \textit{Modern Quantum Mechanics} (Cambridge University Press, 2017).

\bibitem{Shankar} Shankar, R. \textit{Principles of Quantum Mechanics} (Springer, 1994).

\bibitem{Cohen-Tannoudji} Cohen-Tannoudji, C., Dupont-Roc, J., \& Grynberg, G. \textit{Photons and Atoms: Introduction to Quantum Electrodynamics} (Wiley-VCH, 2004).

\bibitem{Kosmann-Schwarzbach} Kosmann-Schwarzbach, Y. \textit{The Noether Theorems: Invariance and Conservation Laws in the Twentieth Century} (Springer, 2011).

\bibitem{Banados2016} Banados, M. \& Reyes, I. A short review on Noether's theorems, gauge symmetries and boundary terms. \href{https://doi.org/10.1142/S0218271816300214}{\textit{Int. J. Mod. Phys. D} \textbf{25}, 1630021 (2016)}.

\bibitem{Fushchich} Fushchich, W. I. \& Nikitin, A. G. \textit{Symmetries of Maxwell's Equations. Mathematics and its Applications} (Springer, 1987).

\bibitem{Pozar} Pozar, D. M. \textit{Microwave Engineering: Theory and Techniques} (John Wiley \& Sons, 2021).

\bibitem{Ziolkowski2001} Ziolkowski, R. W. \& Heyman, E. Wave propagation in media having negative permittivity and permeability. \href{https://doi.org/10.1103/PhysRevE.64.056625}{\textit{Phys. Rev. E} \textbf{64}, 056625 (2001)}.

\bibitem{Veselago2006} Veselago, V. G. \& Narimanov, E. E. The left hand of brightness: past, present and future of negative index materials. \href{https://doi.org/10.1038/nmat1746}{\textit{Nat. Mater.} \textbf{5}, 759 (2006)}.

\bibitem{Pendry2004} Pendry, J. B. Negative refraction. \href{https://doi.org/10.1080/00107510410001667434}{\textit{Contemp. Phys.} \textbf{45}, 191 (2004)}.

\bibitem{Pendry2008} Pendry, J. B. Time reversal and negative refraction. \href{https://doi.org/10.1126/science.1162087}{\textit{Science} \textbf{322}, 71 (2008)}.

\bibitem{Caloz} Caloz, C. \& Itoh, T. \textit{Electromagnetic Metamaterials: Transmission Line Theory and Microwave Applications} (John Wiley \& Sons, 2005).

\bibitem{Engheta} Engheta, N. \& Ziolkowski, R. W. \textit{Metamaterials: Physics and Engineering Explorations} (John Wiley \& Sons, 2006).

\bibitem{Foster1924} Foster, R. M. A reactance theorem. \href{https://doi.org/10.1002/j.1538-7305.1924.tb01358.x}{\textit{Bell Syst. Tech. J.} \textbf{3}, 259 (1924)}.

\bibitem{Dolling2006} Dolling, G., Enkrich, C., Wegener, M., Soukoulis, C. M., \& Linden, S. Simultaneous negative phase and group velocity of light in a metamaterial. \href{https://doi.org/10.1126/science.1126021}{\textit{Science} \textbf{312}, 892 (2006)}.

\bibitem{Ugarte-Munoz2012} Ugarte-Mu\~noz, E., Hrabar, S., Segovia-Vargas, D., \& Kiricenko, A. Stability of non-Foster reactive elements for use in active metamaterials and antennas. \href{https://doi.org/10.1109/TAP.2012.2196957}{\textit{IEEE Trans. Antennas Propag.} \textbf{60}, 3490 (2012)}.

\bibitem{Saadat2012} Saadat, S., Adnan, M., Mosallaei, H., \& Afshari, E. Composite metamaterial and metasurface integrated with non-Foster active circuit elements: A bandwidth-enhancement investigation. \href{https://doi.org/10.1109/TAP.2012.2227654}{\textit{IEEE Trans. Antennas Propag.} \textbf{61}, 1210 (2012)}.

\bibitem{Linvill1953} Linvill, J. G. Transistor negative-impedance converters. \href{https://doi.org/10.1109/JRPROC.1953.274251}{\textit{Proc. IRE} \textbf{41}, 725 (1953)}.

\bibitem{Gonzalez-Posadas2010} Gonz\'alez-Posadas, V., Segovia-Vargas, D., Ugarte-Mu\~noz, E., Jim\'enez-Mart\'in, J. L., \& Garc\'ia-Mu\~noz, L. E. On the performance of negative impedance converters (NICs) to achieve active metamaterials, in \href{https://ieeexplore.ieee.org/document/5729686}{\textit{20th International Conference on Applied Electromagnetics and Communications (ICECom)} (IEEE, 2010) pp. 1-4}.

\bibitem{Hrabar2010B} Hrabar, S., Krois, I., \& Kiricenko, A. Towards active dispersionless ENZ metamaterials for cloaking applications. \href{https://doi.org/10.1016/j.metmat.2010.07.001}{\textit{Metamaterials} \textbf{4}, 89 (2010)}.

\bibitem{Barbuto2013} Barbuto, M., Monti, A., Bilotti, F., \& Toscano, A. Design of a non-Foster actively loaded SRR and application in metamaterial-inspired components. \href{https://doi.org/10.1109/TAP.2012.2228621}{\textit{IEEE Trans. Antennas Propag.} \textbf{61}, 1219 (2013)}.

\bibitem{Hrabar2013A} Hrabar, S., Krois, I., Bonic, I., \& Kiricenko, A. Ultra-broadband simultaneous superluminal phase and group velocities in non-Foster epsilon-near-zero metamaterial. \href{https://doi.org/10.1063/1.4790297}{\textit{Appl. Phys. Lett.} \textbf{102}, 054108 (2013)}.

\bibitem{Hrabar2011} Hrabar, S., Krois, I., Bonic, I., \& Kiricenko, A. Negative capacitor paves the way to ultra-broadband metamaterials. \href{https://doi.org/10.1063/1.3671366}{\textit{Appl. Phys. Lett.} \textbf{99}, 254103 (2011)}.

\bibitem{Okorn2017} Okorn, B., Hrabar, S., \& Krois, I. Physically sound model of a non-Foster negative capacitor. \href{https://doi.org/10.1080/00051144.2017.1398211}{\textit{Automatika} \textbf{58}, 244 (2017)}.

\bibitem{Zanic2021} Zanic, D., Lebo, K., Hrabar, S. Open circuit stable non-Foster negative inductor, in \href{https://doi.org/10.1109/Metamaterials52332.2021.9577184}{\textit{15th International Congress on Artificial Materials for Novel Wave Phenomena (Metamaterials)}, (IEEE, 2021) pp. 21-23}.

\bibitem{Albarracin-Vargas2016} Albarrac\'in-Vargas, F., Gonz\'alez-Posadas, V., Herra\'iz-Martinez, F. J., \& Segovia-Vargas, D. Design method for actively matched antennas with non-Foster elements. \href{https://doi.org/10.1109/TAP.2016.2583482}{\textit{IEEE Trans. Antennas Propag.} \textbf{64}, 4118 (2016)}.

\bibitem{Albarracin-Vargas2023} Albarrac\'in-Vargas, F., Gonz\'alez-Posadas, V., Segovia-Vargas, D. Small printed antenna array based on non-Foster networks, in \href{https://doi.org/10.1109/Metamaterials58257.2023.10289315}{\textit{17th International Congress on Artificial Materials for Novel Wave Phenomena (Metamaterials)}, (IEEE, 2023) pp. 332-334}.

\bibitem{Wang2024} Wang, S. \& Li, Y. Single-transistor impedance matching circuit for over hundred-octave active antennas. \href{https://doi.org/10.1109/TAP.2024.3351203}{\textit{IEEE Trans. Antennas Propag.} \textbf{72}, 2391 (2024)}.

\bibitem{Vehmas2014} Vehmas, J., Hrabar, S., \& Tretyakov, S. A. Transmission lines emulating moving media. \href{https://doi.org/10.1088/1367-2630/16/9/093065}{\textit{New J. Phys.} \textbf{16}, 093065 (2014)}.

\bibitem{Hrabar2013B} Hrabar, S., Krois, I., Bonic, I., Kiricenko, A., \& Muha, D. Active reconfigurable metamaterial unit cell based on non-Foster elements, \href{https://doi.org/10.21236/ada591562}{Final Report for Contract FA8655-12-1-2081, EOARD/AFRL (2013)}.

\bibitem{Qin2023} Qin, X., Fu, P., Yan, W., Wang, S., Lv, Q., \& Li, Y. Negative capacitors and inductors enabling wideband waveguide metatronics. \href{https://doi.org/10.1038/s41467-023-42808-z}{\textit{Nat. Commun.} \textbf{14}, 7041 (2023)}.

\bibitem{Hrabar2018} Hrabar, S. First ten years of active metamaterial structures with ``negative'' elements. \href{https://doi.org/10.1051/epjam/2018005}{\textit{EPJ Appl. Metamat.} \textbf{5}, 9 (2018)}.

\end{thebibliography}}
\end{document}